\documentclass[twocolumn,showpacs,prd,floatfix,axodraw]{revtex4}
\usepackage{mathrsfs}
\usepackage{graphicx,,booktabs,bm}
\usepackage{overpic}
\usepackage{color}
\usepackage{amssymb}

\usepackage{mathrsfs,bm,amsmath,amssymb}
\usepackage{longtable,lscape}
\usepackage{txfonts}
\usepackage{amssymb}
\usepackage{indentfirst}
\usepackage{graphicx,,booktabs}
\usepackage{multirow,ulem}
\newcommand{\pararrow}{\mathord{\buildrel{\lower3pt\hbox{$\scriptscriptstyle\leftrightarrow$}}\over {\partial}}} 
\newcommand{\pararrowk}[1]{\mathord{\buildrel{\lower3pt\hbox{$\scriptscriptstyle\leftrightarrow$}}\over {\partial}\hspace*{-0.18em}{}^#1}\hspace*{-0.18em} } 

\usepackage{color}
\usepackage{amssymb}

\definecolor{cover}{rgb}{0.77,0.87,0.88}
\definecolor{blueone}{rgb}{0.1,0.1,.7}
\definecolor{citec}{rgb}{0.14,0.47,0.09}
\definecolor{two}{rgb}{0.0,0.5,0.}
\definecolor{three}{rgb}{.5,.1,0.15}
\usepackage[bookmarks=true,bookmarksopen=false,plainpages=false,breaklinks=true,
   bookmarksnumbered=true,hypertexnames=false,
   filecolor=blue,urlcolor=three,menucolor=three,
   linkcolor=three,citecolor=blueone, colorlinks,
   anchorcolor=blue,runcolor=pink,frenchlinks=red
   pdfstartview=FitH,pdftitle=title,%
   pdfauthor=author]{hyperref}

\begin{document}
\title{Interpretation of $\Upsilon(11020)$ as an $S$-Wave $B_1\bar{B}$--$B_1\bar{B}^*$ Molecular State}

\author{Qing Lu}
\affiliation{Key Laboratory of Computational Physics of Sichuan Province, School of Mathematics and Physics, Yibin University, Yibin 644000, China}

\author{Cai Cheng}
\affiliation{School of Physics and Electronic Engineering, Sichuan Normal University, Chengdu 610101, China}

\author{Yin Huang\footnote{corresponding author}} \email{huangy2019@swjtu.edu.cn}
\affiliation{School of Physical Science and Technology, Southwest Jiaotong University, Chengdu 610031,China}

\begin{abstract}
Although heavy-quark symmetry implies a $B_1\bar{B}$ molecular partner of the $D_1\bar{D}$ molecule, no such state has been observed experimentally to date. In this work, we propose that the experimentally observed $\Upsilon(11020)$ may be a candidate for such a molecular state, although it may also contain a $B_1\bar{B}^{*}$ component. To explore whether the $\Upsilon(11020)$ can be interpreted as an $S$-wave $B_1\bar{B}$--$B_1\bar{B}^{*}$ molecule, we calculate its strong decay widths within this molecular scenario using the compositeness condition and effective Lagrangians. The coupling of the $\Upsilon(11020)$ to its constituents $B_1$ and $\bar{B}^{(*)}$ is determined by fitting the available experimental data on $\Upsilon(11020) \to e^+ e^-$ and $\Upsilon(11020) \to \chi_{bJ} \pi\pi\pi$ decays.
With the extracted coupling, we compute the partial decay widths of the $\Upsilon(11020)$ into $B^{(*)}_{(s)}\bar{B}^{(*)}_{(s)}$, $\pi\pi \Upsilon(nS)$, $\pi\pi h_b(nP)$, and $\pi\pi\pi \chi_{b1}$ via hadronic loops, as well as the three-body $B^{*}\pi \bar{B}^{(*)}$ decays via tree-level diagrams. Our results suggest that the $\Upsilon(11020)$ can indeed be interpreted as an $S$-wave $B_1\bar{B}$--$B_1\bar{B}^{*}$ molecule with a dominant $B_1\bar{B}$ component, whose main decay channel is $B_s^{*}\bar{B}^{*}$. Moreover, the partial widths for $\Upsilon(11020)\to \pi\pi \Upsilon(nS)$ and $\Upsilon(11020)\to \pi\pi h_b(nP)$ are found to be only a few eV. In contrast, the widths for $\Upsilon(11020)\to \pi\pi\pi \chi_{bJ}$ are considerably larger, with $\pi\pi\pi \chi_{b1}$ reaching up to 0.167~MeV, while the yet-unobserved $\pi\pi\pi \chi_{b0}$ channel could be as large as 0.754~keV.
These distinctive decay patterns could serve as experimental signatures of the molecular nature of the $\Upsilon(11020)$, and their confirmation would provide a test of the applicability of heavy-quark symmetry.
\end{abstract}

\date{\today}


\maketitle
\section{Introduction}\label{sec:intro}
With the progress of experimental studies, several hundred hadrons have been discovered to date~\cite{ParticleDataGroup:2024cfk}.
Among them, many can be interpreted within the conventional quark model as mesons composed of a quark--antiquark pair
and baryons consisting of three quarks~\cite{Godfrey:1985xj,Koniuk:1979vy,Isgur:1978wd}. However, a growing number of hadrons exhibit
more intricate internal structures, commonly referred to as exotic states.  The first established example, the heavy-quark state $X(3872)$,
was observed in 2003 in the $B^{\pm} \to K^{\pm}\pi^{+}\pi^{-}J/\psi$ decay through the $\pi^{+}\pi^{-}J/\psi$ invariant-mass
spectrum~\cite{Belle:2003nnu}, pointing to a structure beyond the conventional $q\bar q$ picture, involving at least four quarks.
This discovery, together with the subsequent observation of many other exotic candidates, has raised fundamental questions about the
spectrum of hadronic structures allowed by Quantum Chromodynamics (QCD).

QCD permits hadrons to realize a variety of configurations as long as they satisfy the color-singlet condition,
including compact multiquark states, hadronic molecules, glueballs, and quark--gluon hybrids~\cite{Chen:2022asf,Klempt:2007cp,Guo:2017jvc}.
Yet quark confinement and asymptotic freedom make their internal structures notoriously difficult to pin down.
Consequently, uncovering the configurations of these exotic hadrons has become a central challenge in contemporary hadron physics.
Among the possible interpretations, the hadronic molecular picture has emerged as an important paradigm, offering a framework to interpret and
predict exotic states whose internal dynamics go beyond conventional quark configurations~\cite{Godfrey:1985xj,Koniuk:1979vy,Isgur:1978wd}.
This perspective is firmly motivated by the fact that hadronic molecules exist in nature, with the deuteron---a weakly bound proton--neutron
system stabilized by the nuclear force---serving as the prototype example.

A number of observed hadrons can indeed be interpreted as molecular states, among which $\Lambda(1405)$ is the most prominent
example in the light-quark sector. Both phenomenological models~\cite{Oset:1997it,Oller:2000fj,Miyahara:2015bya,Kamiya:2016jqc} and lattice
QCD calculations~\cite{Nemoto:2003ft,Hall:2014uca} indicate that it is predominantly a $\bar{K}N$ molecular state.  In the heavy-quark sector,
numerous candidates for molecular states have been reported. The experimentally observed $X(3872)$~\cite{Belle:2003nnu}, with a mass very close to
the $D\bar{D}^{*}$ threshold, has been widely studied as a $D\bar{D}^{*}$ molecular candidate~\cite{Guo:2017jvc,Suzuki:2005ha,Wong:2003xk,AlFiky:2005jd}.
The $D_{s0}(2317)$ and $D_{s1}(2460)$ exhibit anomalously low masses, lying about 160 MeV and 70 MeV below quark-model predictions~\cite{Godfrey:1985xj},
respectively.  These anomalies can be naturally explained as $KD$ and $KD^{*}$ molecular states~\cite{BaBar:2003oey,CLEO:2003ggt,Belle:2003kup,Belle:2003guh,BaBar:2004yux,Meng:2022ozq,Xie:2010zza,Guo:2006fu,Guo:2006rp,Gamermann:2006nm,
Zhu:2019vnr,Mohler:2013rwa,Altenbuchinger:2013vwa}, thus strongly supporting their interpretation as molecular candidates.  Furthermore, the LHCb
Collaboration has reported several hidden-charm pentaquark states~\cite{LHCb:2019kea,LHCb:2015yax,LHCb:2016ztz,LHCb:2016lve} that exhibit features
consistent with $\Sigma_c\bar{D}^{(*)}$ molecules~\cite{Chen:2019bip,Guo:2019fdo,Xiao:2019aya,He:2019ify,Xiao:2019mvs,Roca:2015dva,Chen:2015moa,Chen:2015loa,Yang:2015bmv,Huang:2015uda}.

In addition to the experimentally observed molecular candidates, many predicted states remain unconfirmed, posing significant challenges to existing
theoretical frameworks. A particularly relevant framework is heavy-quark symmetry (HQS)~\cite{Isgur:1991wq}, which refers to the approximate invariance
of the strong interaction under changes in the spin and flavor of a heavy quark when its mass is much larger than the QCD scale.  In particular, heavy-quark
spin symmetry (HQSS) predicts a $J^{PC} = 2^{++}$ partner of the $X(3872)$ with a $D^*\bar{D}^*$ molecular component~\cite{Guo:2013sya}.  The Belle Collaboration
reported evidence for such a state in 2022 via $\gamma\gamma \to \gamma \psi(2S)$, with a global significance of only $2.8\sigma$~\cite{Belle:2021nuv},
highlighting the need for further investigation.  Similarly, no molecular partners of $K D$ and $K D^*$---namely $K \bar{B}$ and $K \bar{B}^*$---or of
$\Sigma_c \bar{D}^{(*)}$---the hidden-bottom pentaquark molecules $\Sigma_B B^{(*)}$---have yet been observed~\cite{ParticleDataGroup:2024cfk}.

Uncovering the heavy-quark-flavor partner candidates of the $D_1\bar{D}$ molecular state is the central objective of this work.  Experimentally, two states
have been identified that can be interpreted as $D_1\bar{D}$ molecular candidates~\cite{Ding:2008gr,Liu:2013vfa,Cleven:2013mka,Qin:2016spb,Cleven:2016qbn,Wang:2020lua,Lee:2008tz,Wang:2013cya},
namely $Z_2^+(4250)$~\cite{Belle:2008qeq} and $Y(4260)$~\cite{BaBar:2005hhc}.  Its heavy-quark-flavor symmetry partner, $B_1\bar{B}$, is generally regarded
as unobserved; we argue, however, that it corresponds to the experimentally established $\Upsilon(11020)$.  The $\Upsilon(11020)$, the highest-mass $\Upsilon$
state known to date~\cite{ParticleDataGroup:2024cfk}, was first reported by the CUSB experiment in 1985~\cite{Lovelock:1985nb} and soon confirmed by the CLEO Collaboration~\cite{CLEO:1984vfn}.
Its existence was subsequently confirmed with higher precision by the BaBar and Belle experiments in 2008~\cite{BaBar:2008cmq}, 2015~\cite{Belle:2015aea,Belle:2015tbu},
and 2019~\cite{Belle:2019cbt} through various production channels.  According to the most recent PDG compilation~\cite{ParticleDataGroup:2024cfk}, its reported mass, width,
and spin--parity are
\begin{align}
 M &= 11000 \pm 4~\mathrm{MeV},~
 \Gamma &= 24^{+8}_{-6}~\mathrm{MeV},~
 I(J^{PC}) &= 0~(1^{--}). \nonumber
\end{align}

Currently, $\Upsilon(11020)$ is commonly identified as $\Upsilon(6S)$, an $S$-wave vector bottomonium state. However, its decay properties deviate from the conventional quarkonium picture.
Experiments show that $\Upsilon(nS)\pi\pi$ and $h_b(kP)\pi\pi$ production at the $\Upsilon(6S)$ occurs almost entirely through intermediate $Z_b^{(')}$ (corresponding to $Z_b(10610)$ and
$Z_b(10650)$) states~\cite{Belle:2015aea,Belle:2015tbu}, leaving negligible nonresonant contributions. By contrast, at $\Upsilon(5S)$ a substantial fraction of $\Upsilon(nS)\pi\pi$ yield
occurs outside the $Z_b^{(')}$ channel. This striking difference suggests a structural origin for $\Upsilon(6S)$ distinct from its lower-mass sibling.

Given the prevailing interpretations that $Z_b$ and $Z_b^{'}$ correspond to $B\bar{B}^{*}$ and $B^{*}\bar{B}^{*}$ molecular states, respectively, and the observation that
$\Upsilon(6S) \to \Upsilon(nS)\pi\pi$ and $\Upsilon(6S) \to h_b(kP)\pi\pi$ proceed via $Z_b^{(*)}$ intermediates, we propose that $\Upsilon(11020)$ may have a molecular configuration
composed of $B_1\bar{B}$ and $B_1\bar{B}^{*}$. In this picture, decays are governed by hadronic degrees of freedom, with the dominant mechanism $B_1\bar{B}^{(*)} \to (B^{*}\bar B^{(*)}/\bar{B}^{*} B^{(*)}) + \pi \to Z_b^{(*)} + \pi$, providing a common pathway to both HQSS-violating ($h_b\pi\pi$) and HQSS-conserving ($\Upsilon(nS)\pi\pi$) channels. This naturally explains the
$Z_b^{(*)}$-dominated decay pattern and the near absence of nonresonant contributions, with HQSS breaking arising from the light-quark spin--orbit structure of $B_1$. Thus, $\Upsilon(6S)$
serves as a critical platform for exploring exotic hadronic molecules.

To assess whether $\Upsilon(11020)$ can be interpreted as a $B_1\bar{B}^{(*)}$ molecular state, one can consider its proximity to the $B_1\bar{B}^{(*)}$ threshold, complemented by calculations
of interactions and decay properties. Its central mass ($M = 11000~\text{MeV}$) lies only $5.41~\text{MeV}$ below the $B_1\bar{B}$ threshold, consistent with a yet-unobserved $B_1\bar{B}$
molecule, regarded as the heavy-quark-symmetry partner of the $D_1\bar{D}$ molecule. Since $B$ and $B^{*}$ form a heavy-quark-symmetry doublet, the $B_1\bar{B}^{*}$ component is also expected
to exist. Within heavy meson chiral perturbation theory, based on spin--flavor symmetry, calculations show that $B_1$ and $\bar{B}^{(*)}$ interactions can form a molecular state~\cite{Ding:2008gr}.
Using this molecular framework, we evaluate decay widths, particularly strong decays, and compare them with experimental data to test the molecular interpretation.

This paper is organized as follows. In Sec.~\ref{Sec: formulism},  we will present the theoretical  formalism.  In Sec.~\ref{Sec: results}, the numerical result will be given,
followed by discussions and conclusions in the last section.

\section{FORMALISM AND INGREDIENTS}\label{Sec: formulism}
\begin{figure}[http]
\begin{center}
\includegraphics[bb=60 350 1150 710, clip, scale=0.46]{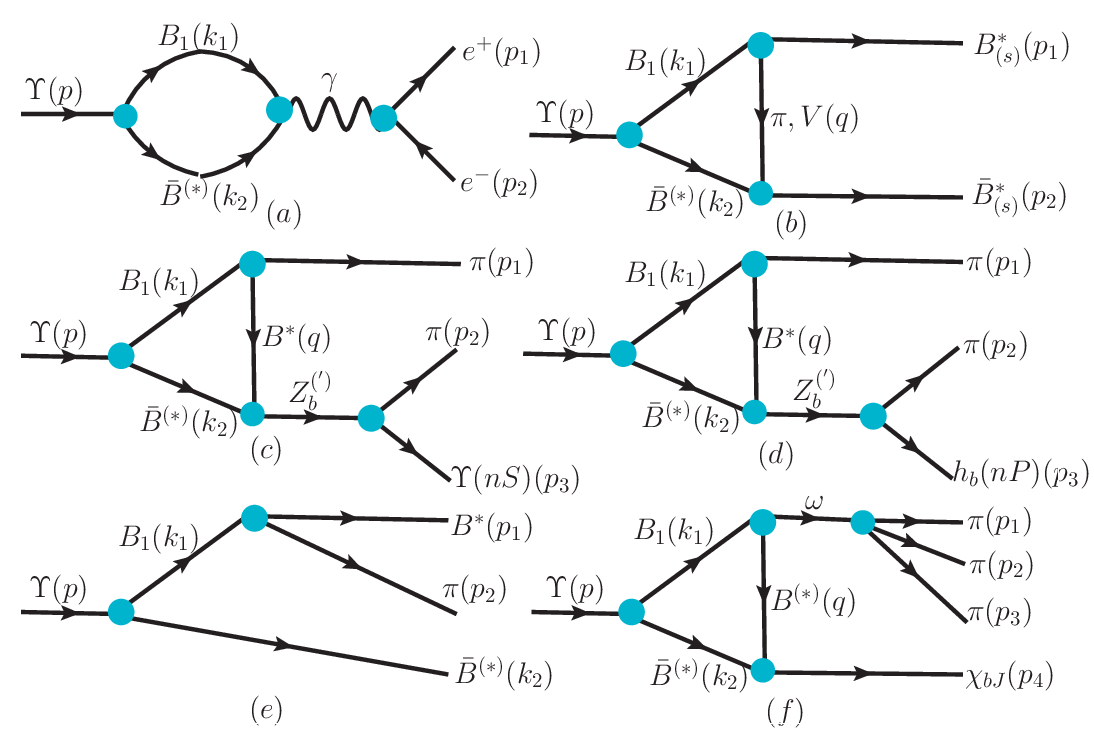}
\caption{Feynman diagrams for the $\Upsilon(11020)$ decays within the $B_1\bar{B}^{(*)}$ molecular framework.
$\Upsilon(nS)$ denotes the $\Upsilon(1S)$, $\Upsilon(2S)$, or $\Upsilon(3S)$ state, while $h_{b}(nP)$ represents the $h_{b}(1P)$ or $h_{b}(2P)$ state.
Similarly, $\chi_{bJ}$ refers to the $\chi_{b0}$, $\chi_{b1}$, and $\chi_{b2}$ states corresponding to $J = 0, 1,$ and $2$, respectively.
The definitions of the kinematic variables ($p$, $p_1$, $p_2$, $p_3$, $p_4$, $q$, $k_1$, and $k_2$) used in the calculation are also indicated.}\label{cc1}
\end{center}
\end{figure}
In this study, we explore whether the $\Upsilon(11020)$ resonance can be interpreted as an $S$-wave $B_1\bar{B}^{(*)}$ molecular configurations through an analysis
of its strong decay behavior.  The relevant Feynman diagrams are shown in Fig.~(\ref{cc1}). To evaluate the Feynman diagrams shown in Fig.~(\ref{cc1}), one first needs
to construct the effective Lagrangian densities associated with the relevant interaction vertices.
For the $\Upsilon(11020)B_1\bar{B}^{(*)}$ vertex, we employ the interaction Lagrangian given in Refs.~\cite{Yang:2022rck,Yue:2024bvy}, expressed as
\begin{align}
\mathcal{L}_{\Upsilon(11020)B_1\bar{B}^{(*)}}&= \int d^4y\Phi(y^2)[g_{\Upsilon(11020)B_1\bar{B}}\Upsilon^{\mu}(x)\bar{B}(x+\omega_{B_1}y)\nonumber\\
&\times\,B_{1\mu}(x-\omega_{\bar{B}}y)+ g_{\Upsilon(11020)B_1\bar{B}^{*}}\epsilon_{\mu\nu\alpha\beta}\partial^{\mu}\Upsilon^{\nu}(x)\nonumber\\
    &\times\bar{B}^{*\alpha}(x+\omega_{B_1}y)B_1^{\beta}(x-\omega_{\bar{B}^{*}}y)]+c.c.,\label{eq1}
\end{align}
where the coefficients $\omega_{B_1}=m_{B_1}/(m_{B_1}+m_{\bar{B}^{(*)}})$ and $\omega_{\bar{B}^{(*)}}=m_{\bar{B}^{(*)}}/(m_{B_1}+m_{\bar{B}^{(*)}})$ depend on the respective masses of the $B_1$ and $\bar{B}^{(*)}$ mesons.  The function $\Phi(y^2)$ represents an effective correlation function that characterizes the spatial distribution of the constituent mesons inside the molecular configuration.
In addition to encoding this internal structure, it also acts as a natural regulator, ensuring that loop integrals remain ultraviolet finite.
Although its explicit form is not unique, it must decrease rapidly in the ultraviolet domain.
Following the conventional treatment in hadronic molecular studies, we adopt a Gaussian-type correlation function whose Fourier transform takes the form
\begin{align}
\Phi(p_E^2/\Lambda^2) = \exp(-p_E^2/\Lambda^2),
\end{align}
where $p_E$ denotes the Euclidean Jacobi momentum, and $\Lambda$ is a size parameter, typically of order 1~GeV, that may vary across different systems.
Here, $\Lambda$ is to be determined through a fit to the experimental data, with the detailed procedure described in the next section.

The coupling constant $g_{\Upsilon(11020)B_1\bar{B}^{(*)}}$ in Eq.~\ref{eq1} is determined using the compositeness condition~\cite{Weinberg:1962hj,Salam:1962ap}.
This condition requires the renormalization constant of the hadronic molecular wave function to vanish,
\begin{align}
Z_{\Upsilon(11020)}= X_{B_1\bar{B}} + X_{B_1\bar{B}^{*}} - \frac{\partial \Sigma^{\perp}_{\Upsilon(11020)}(p)}{(\partial p^2)}|_{p^2 = m^2_{\Upsilon(11020)}}= 0,
\end{align}
where $p$ denotes the four-momentum of the $\Upsilon(11020)$.  $X_{B_1\bar{B}}$ and $X_{B_1\bar{B}^{*}}$ represent the probabilities of finding the $B_1\bar{B}$ and $B_1\bar{B}^{*}$
molecular components in the $\Upsilon(11020)$, respectively, and they satisfy $X_{B_1\bar{B}} + X_{B_1\bar{B}^{*}} = 1$.

\begin{figure}[h!]
\begin{center}
\includegraphics[bb=70 600 950 720, clip, scale=0.54]{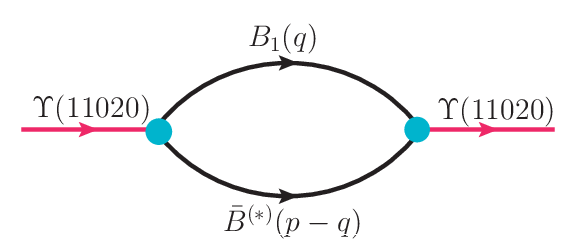}
\caption{Mass operator of the $\Upsilon(11020)$, interpreted as a $B_1\bar{B}^{(*)}$ molecular state.
Here, $q$ and $p$ denote the four-momenta of the $B_1$ meson and the initial $\Upsilon(11020)$, respectively.}
\label{cc2}
\end{center}
\end{figure}
The transverse component $\Sigma^{\perp}_{\Upsilon(11020)}(p)$ of the mass operator $\Sigma^{\mu\nu}_{\Upsilon(11020)}(p)$ is related by
\begin{align}
\Sigma^{\mu\nu}_{\Upsilon(11020)}(p)= (g^{\mu\nu} - p^{\mu}p^{\nu}/p^2)\Sigma^{\perp}_{\Upsilon(11020)}(p) + \cdots.
\end{align}
Using the effective Lagrangian in Eq.~\ref{eq1}, the mass operator $\Sigma^{\mu\nu}_{\Upsilon(11020)}(p)$ corresponding to Fig.~\ref{cc2}
can be expressed as
\begin{align}
\Sigma&^{\mu\nu}_{\Upsilon(11020)}(p)= \Sigma^{\mu\nu}_{\Upsilon(11020)B_1\bar{B}}(p)+\Sigma^{\mu\nu}_{\Upsilon(11020)B_1\bar{B}^{*}}(p)=\int \frac{d^4q}{(2\pi)^4}\nonumber\\
                &\times\Phi^2(q-\omega_{B_1}p)\frac{-g^{ij} + q^{i}q^{j}/m^2_{B_1}}{q^2 -m^2_{B_1}}[\frac{-g^2_{\Upsilon(11020)B_1\bar{B}} }{(p-q)^2- m^2_{\bar{B}}}-\epsilon_{\delta\mu\alpha\beta}\nonumber\\
                 &\times p^{\delta}\epsilon_{\zeta\nu\eta\lambda}p^{\zeta}\frac{-g^{\alpha\eta}+(p-q)^{\alpha}(p-q)^{\eta}/m^2_{\bar{B}^{*}}}{(p-q)^2-m^2_{\bar{B}^{*}}}g^2_{\Upsilon(11020)B_1\bar{B}^{*}}],
\end{align}
where $\Sigma^{\mu\nu}_{\Upsilon(11020)B_1\bar{B}}(p)$ and $\Sigma^{\mu\nu}_{\Upsilon(11020)B_1\bar{B}^{*}}(p)$ denote the self-energy contributions from the $B_1\bar{B}$ and $B_1\bar{B}^{*}$
molecular components, with the indices $ij = \mu\nu$ and $\beta\lambda$ corresponding to the $B_1\bar{B}$ and $B_1\bar{B}^{*}$ systems, respectively.  One should note that $\omega_{B_1}$ differs
between the $B_1\bar{B}$ and $B_1\bar{B}^{*}$ channels.

To evaluate the Feynman diagrams shown in Fig.~\ref{cc1}, it is necessary to specify the effective Lagrangian densities corresponding to the relevant interaction vertices.
For the \( B_1 B^* \pi \) interaction, we adopt the following Lagrangian, constructed within the framework of chiral perturbation theory (CPT)~\cite{Falk:1992cx}:
\begin{align}
{\cal L}_{TH} &= \frac{g_{TH}}{\Lambda_{\chi}} \mathrm{Tr} \Big\{ \bar{H}_a T^\mu_b \big( i D_\mu {\cal A}\!\!\!/ + i D\!\!\!/ \mathcal{A}_\mu \big)_{ba} \gamma_5 \Big\} + \mathrm{H.c.},\label{eq8}
\end{align}
where \( H \) and \( T \) are the heavy-meson fields, defined as
\begin{align}
H_a &= \frac{1 + v\!\!\!/}{2}\{B^{*}_{a\mu}\gamma^{\mu}-B_{a}\gamma_5\},~~~\bar{H}_a=\gamma^0H^{\dagger}_a\gamma^0,\\
T_a &=\frac{1 + v\!\!\!/}{2}\Big\{B^{*\mu\nu}_{2a}\gamma_{\nu}-B_{1a\nu}\sqrt{\frac{3}{2}}\gamma_5\Big[g^{\mu\nu}-\frac{\gamma^{\nu}(\gamma^{\mu}-v^{\mu})}{3}\Big]\Big\},
\end{align}
with indices \( a \) and \( b \) labeling SU(3) flavor components \((u, d, s)\).  The velocity four-vector of the heavy quark is taken as \( v=(1,\vec{0}) \), and the ratio \( g_{TH}/\Lambda_{\chi} \) in Eq.~\eqref{eq8} serves as an effective coupling constant, where \( \Lambda_{\chi} \) denotes a momentum scale characterizing the convergence of the derivative expansion, typically \( \Lambda_{\chi} \sim 1~\mathrm{GeV} \)~\cite{Falk:1992cx}.   The ratio \( g_{TH}/\Lambda_{\chi} \) is determined from the experimental total decay width of the $B_1(5721)$ state~\cite{ParticleDataGroup:2024cfk}.  The detailed results will be presented in the next section.  Note that the total decay width of $B_1(5721)$ is given by the sum of its partial widths for the $\pi B^{*}$ and $\gamma B^{(*)}$ decay modes.

The remaining quantities appearing in Eq.~\ref{eq8} that have not yet been specified are ${\cal A}_{\mu} = \frac{1}{2}\left(\xi \, \partial_{\mu} \xi^{\dagger} - \xi^{\dagger} \, \partial_{\mu} \xi\right)$ and $D_{\mu} = \partial_{\mu} + V_{\mu}$ with
$V_{\mu} = \frac{1}{2}\left(\xi \, \partial_{\mu} \xi^{\dagger} + \xi^{\dagger} \, \partial_{\mu} \xi\right)$. The fields are defined as
$\xi = \exp\left[i \frac{{\cal M}}{f_\pi}\right]$ with $ f_\pi = 132~\text{MeV}$ and \({\cal M}\) denotes the octet of pseudoscalar mesons:
\begin{align}
{\cal M}=
\begin{pmatrix}
\frac{1}{\sqrt{2}}\pi^{0} + \frac{1}{\sqrt{6}}\eta & \pi^{+} & K^{+} \\
\pi^{-} & -\frac{1}{\sqrt{2}}\pi^{0} + \frac{1}{\sqrt{6}}\eta & K^0 \\
K^{-} & \bar{K}^{0} & -\frac{2}{\sqrt{6}}\eta
\end{pmatrix}\label{eq13-1}.
\end{align}
Expanding to leading order, one obtains
\begin{align}
{\cal A}_{\mu} &= -\frac{i}{f_\pi} \, \partial_\mu {\cal M}, & V_{\mu} &= 0.
\end{align}

The effective Lagrangians describing the interactions at the
$Z_b^{(')}B\bar{B}^{*}$, $Z_b^{(')}\Upsilon(nS)\pi$, and
$Z_b^{(')}h_{b}(nP)\pi$ vertices are given by~\cite{Wang:2023vkx,Huang:2015xud,Chen:2019gfp,Cleven:2013sq}
\begin{align}
{\cal L}_{Z^{(')}_bB^{(*)}\bar{B}^{*}}&= g_{Z_bB\bar{B}^{*}}Z_b^{\mu}\left( B^{*}_{\mu}B^{\dagger}+BB^{*\dagger}_{\mu}\right) \nonumber\\
                                      &+g_{Z_b^{'}B^{*}\bar{B}^{*}}\epsilon_{\mu\nu\alpha\beta}\partial^{\mu}Z_b^{'\nu}B^{*\alpha}B^{*\dagger\beta}+ H.c.,\label{eq13}\\
{\cal L}_{Z^{(')}_b\Upsilon(nS)\pi} &=\frac{g_{Z^{(')}_b\Upsilon(nS)\pi}}{m_{Z_b}}\partial_{\mu}\Upsilon(nS)_{\nu}\nonumber\\
                                    &\times\left(\partial^{\mu}\pi\,Z_b^{(')\nu}- \partial^{\nu}\pi\,Z_b^{(')\mu}\right) + H.c.,\label{eq14}\\
{\cal L}_{Z^{(')}_bh_{b}(nP)\pi} &=g_{Z^{(')}_bh_{b}(nP)\pi}\epsilon_{\mu\nu\alpha\beta}Z_b^{(')\mu}\partial^{\nu}h_{b}(nP)^{\alpha}\partial^{\beta}\pi + H.c.,\label{eq15}
\end{align}
where the coupling constants $g_{Z_bB\bar{B}^{*}}$ and $g_{Z^{'}_bB^{*}\bar{B}^{*}}$ are extracted from the experimentally measured partial widths of $Z_b \to B\bar{B}^{*} + B^{*}\bar{B}$
and $Z_b^{'}\to B^{*}\bar{B}^{*}$, respectively.  Using the Lagrangian in Eq.~(\ref{eq13}), the two-body decay widths $\Gamma(Z_b^{+} \to B^{*+} \bar{B}^0)$ and
$\Gamma(Z_b^{'+} \to B^{*+} \bar{B}^{*0})$ are related to the corresponding coupling constants as
\begin{align}
\Gamma(Z_b^{+} \to B^{*+} \bar{B}^0)&=\frac{g^2_{Z_bB\bar{B}^{*}}}{24\pi m^2_{Z_b}}( \frac{{\cal P}^2_{B^{*}\bar{B}^{0}}}{m^2_{B^{*}}} + 3 ){\cal P}_{B^{*}\bar{B}^{0}}\simeq \frac{1}{2}\,\Gamma_{Z_b^{+}}, \nonumber\\
\Gamma(Z_b^{'+} \to B^{*+} \bar{B}^{*0})&=\frac{g^2_{Z^{'}_bB^{*}\bar{B}^{*}}}{24\pi }(\frac{m^2_{Z^{'}_b}}{m^2_{B^{*}}}+2){\cal P}_{B^{*}\bar{B}^{*0}},\nonumber
\end{align}
where ${\cal P}_{B^{*}\bar{B}^{0}}$ and ${\cal P}_{B^{*}\bar{B}^{*0}}$ denote the three-momenta of the $\bar{B}^0$ and $\bar{B}^{*0}$ in the rest frames of $Z_b$ and $Z_b^{'}$,
respectively.  Using the experimental total widths $\Gamma_{Z_b^{+}} = 18.4 \pm 2.4~\text{MeV}$ and $\Gamma_{Z_b^{'+}} = 11.5 \pm 2.2~\text{MeV}$, together with the
measured branching fractions---accounting for $85.9\%$ and $74\%$ of the total widths, respectively---and the particle masses from Ref.~\cite{ParticleDataGroup:2024cfk},
we obtain $g_{Z_bB\bar{B}^{*}} = 13.52~\text{GeV}$ and $g_{Z^{'}_bB^{*}\bar{B}^{*}} = 0.94~\text{GeV}$.

For the coupling constants $g_{Z^{(')}_b\Upsilon(nS)\pi}$ and $g_{Z^{(')}_bh_b(nP)\pi}$, their values are also determined from experimental data~\cite{ParticleDataGroup:2024cfk}.
The relations between these couplings and the corresponding partial decay widths can be derived from Eqs.~(\ref{eq14}) and (\ref{eq15}), and are given by
\begin{align}
\Gamma(Z_b^{(')+}&\to{}\Upsilon(nS)\pi)=\frac{g^2_{Z_b^{(')}\Upsilon(nS)\pi}}{96\pi m^6_{Z^{(')}_b}}{\cal P}_{\Upsilon_n\pi}\Big[(2m^2_{Z^{(')}_b}+m^2_{\Upsilon_n})\nonumber\\
                                &\times(m^2_{Z^{(')}_b}-m^2_{\Upsilon_n}-m^2_{\pi})^2+8m^2_{Z^{(')}_b}m^2_{\pi}(m^2_{Z^{(')}_b}-m^2_{\Upsilon_n})\Big],\nonumber\\
\Gamma(Z_b^{(')+}&\to{}h_b(nP)\pi)=\frac{g^2_{Z^{'}_bh_b(nP)\pi}}{12\pi}{\cal P}_{h_b\pi}^3,\nonumber
\end{align}
where ${\cal P}_{\Upsilon/h_b\pi}$ denotes the three-momentum of the $\pi$ meson in the $Z_b$ rest frame. $m_{Z^{(')}_b}$, $m_{h_b}$, $m_{\Upsilon_n}$, and $m_{\pi}$ represent
the masses of the $Z_b^{(')}$, $h_b(nP)$, $\Upsilon(nS)$, and $\pi$ mesons, respectively.  The coupling constants are extracted directly from the experimental values listed in
Table~\ref{table1}, and the corresponding results are summarized therein.
\begin{table}[h!]
\centering
\tabcolsep=1.2mm
\caption{Experimental partial decay widths (central values) $\Gamma$ (in MeV) for the processes $Z_b^{(')+}\!\to\!\Upsilon(nS)\pi^{+}$ and $Z_b^{(')+}\!\to\!h_b(nP)\pi^{+}$,
together with the corresponding coupling constants $g_{Z^{(')}_b}$.  Here, $g_{Z^{(')}_b\pi\Upsilon}$ is dimensionless, whereas $g_{Z^{(')}_b\pi h_b}$ is expressed in units
of $\mathrm{GeV^{-1}}$.} \label{table1}
\begin{tabular}{cccccc}
\hline\hline
         & $\Upsilon(1S)\pi^{+}$ & $\Upsilon(2S)\pi^{+}$ & $\Upsilon(3S)\pi^{+}$ & $h_b(1P)\pi^{+}$ & $h_b(2P)\pi^{+}$ \\
\hline
$\Gamma$ ($Z_b\to$) & 0.0994 & 0.6624 & 0.3864 & 0.6440 & 0.8648 \\
$g_{Z_b}$ & 0.4876 & 3.3043 & 9.2470 & 0.2836 & 1.0315 \\
$\Gamma$ ($Z^{'}_b\to$) & 0.0196 & 0.1610 & 0.1840 & 0.9660 & 1.7250 \\
$g_{Z^{'}_b}$ & 0.2058 & 1.4670 & 4.9020 & 0.3165 & 1.1807 \\
\hline\hline
\end{tabular}
\end{table}

Based on the heavy-quark limit and chiral symmetry~\cite{Casalbuoni:1996pg}, the interactions between light vector and pseudoscalar mesons and the heavy bottom mesons are expressed as~\cite{Wu:2022hck,Liu:2023gtx,Yue:2024bvy,Huang:2017kkg,Colangelo:2002mj,Veliev:2010gb}
\begin{align}
{\cal{L}}_{B^{(*)}B^{(*)}V/{\cal{} M}}&=-ig_{BBV}B_i^{\dagger}\pararrowk{\mu} B_{j}(V^{\dagger}_{\mu})_{ij}+ig_{B^{*}B^{*}V}B^{*\nu\dagger}_i\pararrowk{\mu}B^{*}_{\nu j}(V^{\dagger}_{\mu})_{ij}\nonumber\\
         &-2f_{B^{*}BV}\epsilon_{\mu\nu\alpha\beta}(\partial^{\mu}V^{\nu\dagger})_{ij}(B^{\dagger}_{i}\pararrowk{\alpha}B^{*\beta}_j-B_i^{*\beta\dagger}\pararrowk{\alpha}B_j)\nonumber\\
         &+i4f_{B^{*}B^{*}V}B_{i\mu}^{*\dagger}(\partial^{\mu}V^{\nu\dagger}-\partial^{\nu}V^{\mu\dagger})_{ij}B_{\nu j}^{*}\nonumber\\
         &-ig_{B^{*}B{\cal{M}}}(B_i^{\dagger}B_{j \mu}^{*}-B_{i \mu}^{*\dagger}B_j)\partial^{\mu}{\cal{M}}^{\dagger}_{ij}\nonumber\\
         &+\frac{1}{2}g_{B^{*}B^{*}{\cal{M}}}\epsilon_{\mu\nu\alpha\beta}B_{i}^{*\mu\dagger}\partial^{\nu}{\cal{M}}^{\dagger}_{ij}\pararrowk{\alpha}B_j^{*\beta}+H.c.,\\
{\cal{L}}_{B^{(*)}B_1V}&=ig_{B^{*}B_1V}\epsilon_{\mu\nu\alpha\beta}[ B^{\mu}_{1b}(\partial^{\alpha}B^{*\nu\dagger}_a)-(\partial^{\alpha}B^{\mu}_{1b})B^{*\nu\dagger}_b]V_{ba}^{\beta}\nonumber\\
                       &+g_{BB_1V}B_bB^{\mu\dagger}_{1a}(V_{\mu})_{ba},\\
{\cal{L}}_{\chi_{bJ}B^{(*)}B^{(*)}}&=-g_{\chi_{b0}BB}\chi_{b0}BB^{\dagger}-g_{\chi_{b0}B^{*}B^{*}}\chi_{b0}B^{*}_{\mu}B^{*\mu\dagger}\nonumber\\
                                   &+ig_{\chi_{b1}BB^{*}}\chi^{\mu}_{b1}(B^{*}_{\mu}B^{\dagger}-BB^{*\dagger}_{\mu})-g_{\chi_{b2}BB}\chi_{b2}^{\mu\nu}\partial_{\mu}B\partial_{\nu}B^{\dagger}\nonumber\\
                                   &-ig_{\chi_{b2}B^{*}B}\epsilon_{\mu\nu\alpha\beta}\partial^{\alpha}\chi_{b2}^{\mu\rho}(\partial_{\rho}B^{*\nu}\partial^{\beta}B^{\dagger}-\partial^{\beta}B\partial_{\rho}B^{*\nu\dagger})\nonumber\\
                                   &+g_{\chi_{b2}B^{*}B^{*}}\chi_{b2}^{\mu\nu}(B^{*}_{\mu}B^{*\dagger}_{\nu}+B^{*}_{\nu}B^{*\dagger}_{\mu}),
\end{align}
where $A\pararrowk{\mu}B = A(\partial^{\mu}B) - (\partial^{\mu}A)B$.  ${\cal M}$ denotes the matrix containing the pion fields, as defined in Eq.~(\ref{eq13-1}), and $V$ represents
the vector-meson nonet in matrix form, given by
\begin{align}
V =
\begin{pmatrix}
\frac{1}{\sqrt{2}}(\rho^{0} + \omega) & \rho^{+} & K^{*+} \\
\rho^{-} & \frac{1}{\sqrt{2}}(-\rho^{0} + \omega) & K^{*0} \\
K^{*-} & \bar{K}^{*0} & \phi
\end{pmatrix}.
\label{eq:Vmatrix}
\end{align}
The bottom meson triplets are defined as $B^{(*)}$ = $(B^{(*)+}$, $B^{(*)0},\, B^{(*)}_s)$ and $B_1$ = $(B_1(5721)^0,\, B_1(5721)^+,\, B_{s1}(5830))$.
From the heavy-quark and chiral limits~\cite{Wu:2022hck,Casalbuoni:1996pg,Huang:2017kkg}, the relevant coupling constants can be written as
$g_{B^{*}B^{*}{\cal M}} = g_{B^{*}B{\cal{M}}}/\sqrt{m_Bm_{B^{*}}}=2g/f_{\pi}$,
$f_{B^{*}BV}=f_{B^{*}B^{*}V}/m_{B^{*}}=\lambda g_V/\sqrt{2}$, and
$g_{BBV}=g_{B^{*}B^{*}V}=\beta g_V/\sqrt{2}$,
where $g = 0.59$, $\lambda = 0.56~\mathrm{GeV^{-1}}$, $\beta = 0.9$, and $g_V = m_{\rho}/f_{\pi}$, with the $\rho$-meson mass given by $m_{\rho} = 775.26~\mathrm{MeV}$.
Similar, one obtains $g_{B B_1 V} = 2 g_{V_1} \xi_1 \sqrt{m_B m_{B_1}}/\sqrt{3}$ and
$g_{B^* B_1 V} = g_{V_1} \xi_1/\sqrt{3}$, with $g_{V_1} = 5.9$ and $\xi_1 = -0.1$~\cite{Wu:2022hck,Casalbuoni:1996pg,Yue:2024bvy,Huang:2017kkg}.  The couplings of $\chi_{bJ}$ to $B^{(*)}\bar{B}^{(*)}$ are determined by a single gauge coupling
$g_1 = -\sqrt{m_{\chi_{b0}}/3}/f_{\chi_{b0}}$,
with the decay constant $f_{\chi_{b0}} = 175 \pm 55$~MeV~\cite{Colangelo:2002mj,Veliev:2010gb},
as explicitly given in Ref.~\cite{Huang:2017kkg}.
\begin{align}
g_{\chi_{b0}BB} &= 2/\sqrt{3} g_1 \sqrt{m_{\chi_{b0}} m_B}, &  g_{\chi_{b0}B^*B^*} &= 2/\sqrt{3} g_1 \sqrt{m_{\chi_{b0}} m_{B^*}}, \nonumber\\
g_{\chi_{b1}BB^*} &= 2 \sqrt{2} g_1 \sqrt{m_{\chi_{b1}} m_B m_{B^*}}, & g_{\chi_{b2}BB} &= 2 g_1 \sqrt{m_{\chi_{b2}} m_B}, \nonumber\\
g_{\chi_{b2}BB^*} &= g_1 m_{\chi_{b2}}/\sqrt{m_B^3 m_{B^*}}, &
g_{\chi_{b2}B^*B^*} &= 4 g_1 \sqrt{m_{\chi_{b2}} m_{B^*}}. \nonumber
\end{align}

Finally, we consider the effective Lagrangians that describe the \( B^{(*)}B_1\gamma \) and \( e^{+}e^{-}\gamma \) coupling vertices.
For the \( e^{+}e^{-}\gamma \) interaction, we employ the commonly used and well-established form,
\begin{align}
{\cal L}_{e^{+}e^{-}\gamma} &= -ie\, \bar{\upsilon}\gamma^{\mu}u\, A_{\mu}, \label{eq71}
\end{align}
where \( A_{\mu} \) denotes the photon field, while \( u \) and \( \upsilon \) are the spinor wave functions of the electron and positron, respectively,
and \( e^2 = 4\pi/137 \).  For the \( B^{(*)}B_1\gamma \) couplings, we adopt the effective Lagrangians from Ref.~\cite{Pullin:2021ebn,Yue:2024bvy},
\begin{align}
{\cal L}_{B_1B\gamma} &= i e g_{B_1B\gamma}\, B_1^{\mu} \partial^{\nu} B\, \partial_{\mu}A_{\nu}, \label{eq7}\\
{\cal L}_{B_1B^{*}\gamma} &= i g_{B^{*}B_1\gamma}\epsilon_{\mu\nu\alpha\beta} B_{1}^{\mu} \pararrowk{\alpha} B^{*\nu} A^{\beta}. \label{eq7-2}
\end{align}
The coupling constants \( g_{\gamma B^{+}_1B^{+}} \) and \( g_{\gamma B^{0}_1B^{0}} \) are determined from the partial
decay widths of \( B^{+}_1 \to B^{+}\gamma \) and \( B^{0}_1 \to B^{0}\gamma \), which are obtained from Eq.~(\ref{eq7}):
\begin{align}
\Gamma(B_1 \to B\gamma) &=\frac{g_{B_1B\gamma}^2 e^2}{192\pi}\,m_B^2 m_{B_1}\left(1 - \frac{m_B^2}{m_{B_1}^2}\right)^3.
\end{align}
Since the width of the \( B_1 \to B\gamma \) decay is not well determined experimentally, we adopt the theoretically predicted partial widths from Ref.~\cite{Pullin:2021ebn},
namely,\(\Gamma(B^{+}_1 \to B^{+}\gamma) = 99.3~\mathrm{keV}\) and \(\Gamma(B^{0}_1 \to B^{0}\gamma) = 26.22~\mathrm{keV}\).
Accordingly, the coupling constants are determined to be \( g_{\gamma B^{+}_1B^{+}} = 0.334~\mathrm{GeV^{-1}} \) and \( g_{\gamma B^{0}_1B^{0}} = 0.172~\mathrm{GeV^{-1}} \).
For the coupling constant \( g_{B^{*}B_1\gamma} \), since there is no available experimental or theoretical information, its value will be determined by fitting to the experimental
data in this work.

By combining all the components, we derive the decay amplitudes summarized below
\begin{align}
{\cal{M}}_{a}&=\frac{ie^2g_{B_1B\gamma}g_{B_1B\Upsilon(11020)}}{p^2}\int\frac{d^4k_1}{(2\pi)^4}\Phi[(k_1\omega_{\bar{B}}-k_2\omega_{B_1})_E^2]\nonumber\\
             &\times{}\bar{\upsilon}(p_1)k\!\!\!/_2u(p_2)p^{\alpha}\frac{-g^{\alpha\eta}+k_1^{\alpha}k_1^{\eta}/m^2_{B_1}}{k_1^2-m^2_{B_1}}\frac{1}{k_2^2-m^2_{\bar{B}}}\epsilon_{\eta}(p),\nonumber\\
             &-\frac{ieg_{B_1B^{*}\gamma}g_{B_1B^{*}\Upsilon(11020)}}{p^2}\int\frac{d^4k_1}{(2\pi)^4}\Phi[(k_1\omega_{\bar{B}^{*}}-k_2\omega_{B_1})_E^2]\nonumber\\
              &\times{}\bar{\upsilon}(p_1)\gamma^{\beta}u(p_2)\epsilon_{\mu\nu\alpha\beta}k_2^{\alpha}\frac{-g^{\mu\lambda}+k_1^{\mu}k_1^{\lambda}/m^2_{B_1}}{k_1^2-m^2_{B_1}}\epsilon_{\tau\eta\sigma\lambda}p^{\tau}\nonumber\\
              &\times\epsilon_{\eta}(p)\frac{-g^{\nu\sigma}+k_2^{\nu}k_2^{\sigma}/m^2_{B^{*}}}{k_2^2-m^2_{B^{*}}}\label{eq23},\\
{\cal{M}}_{b1}&=\frac{i\sqrt{6}g_{TH}g_{\pi{}BB^{*}}g_{\Upsilon(11020)B_1\bar{B}}}{4\Lambda_{\chi}f_{\pi}}\int\frac{d^4q}{(2\pi)^4}\Phi[(k_1\omega_{\bar{B}}-k_2\omega_{B_1})_E^2]\nonumber\\
             &\times{}\epsilon_{\alpha}^{*}(p_1)\gamma^{\alpha}(1+v\!\!\!/)\gamma_5(g^{\mu\nu}-\frac{\gamma^{\nu}(\gamma^{\mu}-v^{\mu})}{3})q_{\mu}q\!\!\!/\gamma_5\epsilon_{\eta}(p)\nonumber\\
             &\times{}\frac{-g^{\nu\eta}+k_1^{\nu}k_1^{\eta}/m^2_{B_1}}{k_1^2-m^2_{B_1}}\frac{1}{k_2^2-m^2_{\bar{B}}}q^{\sigma}\epsilon^{*}_{\sigma}(p_2)\frac{1}{q^2-m^2_{\pi}}\nonumber\\
             &-\frac{\sqrt{6}g_{TH}g_{\pi{}B^{*}B^{*}}g_{\Upsilon(11020)B_1\bar{B}^{*}}}{8\Lambda_{\chi}f_{\pi}}\int\frac{d^4q}{(2\pi)^4}\Phi[(k_1\omega_{\bar{B}^{*}}-k_2\omega_{B_1})_E^2]\nonumber\\
             &\times{}\epsilon_{\sigma}^{*}(p_1)\gamma^{\sigma}(1+v\!\!\!/)\gamma_5(g^{\mu\nu}-\frac{\gamma^{\nu}(\gamma^{\mu}-v^{\mu})}{3})q_{\mu}q\!\!\!/\gamma_5\epsilon_{\eta}(p)\nonumber\\
             &\times\frac{-g^{\nu\beta}+k_1^{\nu}k_1^{\beta}/m^2_{B_1}}{k_1^2-m^2_{B_1}}\epsilon_{\lambda\eta\alpha\beta}p^{\lambda}\frac{-g^{\alpha\tau}+k_2^{\alpha}k_2^{\tau}/m^2_{B^{*}}}{k_2^2-m^2_{B^{*}}}\epsilon_{\tau\varsigma\xi\varphi}\nonumber\\
             &\times{}q^{\varsigma}(p_2^{\xi}+q^{\xi})\epsilon^{*}_{\varphi}(p_2)\frac{1}{q^2-m^2_{\pi}}-\sum_{V=\rho^0,\omega}ig_{B_1B^{*}V}f_{B^{*}BV}\nonumber\\
             &\times{}g_{B_1B\Upsilon(11020)}\int\frac{d^4k_1}{(2\pi)^4}\Phi[(k_1\omega_{\bar{B}}-k_2\omega_{B_1})_E^2]\epsilon_{\mu\nu\alpha\beta}\nonumber
\end{align}
\begin{align}
              &\times{}(p_1+k_1)^{\alpha}\epsilon^{*\nu}(p_1)\frac{-g^{\mu\eta}+k_1^{\mu}k_1^{\eta}/m^2_{B_1}}{k_1^2-m^2_{B_1}}\epsilon_{\eta}(p)\frac{1}{k_2^2-m^2_{\bar{B}}}\nonumber\\
             &\times{}\epsilon_{\sigma\lambda\tau\xi}\epsilon^{\xi}(p_2)q^{\sigma}(k_2+p_2)^{\tau}\frac{-g^{\beta\lambda}+q^{\beta}q^{\lambda}/m_V^2}{q^2-m_V^2}\nonumber\\
              &+\sum_{V=\rho^0,\omega}g_{B_1B^{*}V}g_{B_1B^{*}\Upsilon(11020)}\int\frac{d^4k_1}{(2\pi)^4}\Phi[(k_1\omega_{\bar{B}^{*}}-k_2\omega_{B_1})_E^2]\nonumber\\
             &\times{}\epsilon_{\mu\nu\alpha\beta}(p_1+k_1)^{\alpha}\epsilon^{*\nu}(p_1)\frac{-g^{\mu\sigma}+k_1^{\mu}k_1^{\sigma}/m^2_{B_1}}{k_1^2-m^2_{B_1}}\epsilon_{\tau\eta\lambda\sigma}p^{\tau}\epsilon_{\eta}(p)\nonumber\\
             &\times\frac{-g^{\lambda\xi}+k_2^{\lambda}k_2^{\xi}/m^2_{\bar{B}^{*}}}{k_2^2-m^2_{\bar{B}^{*}}}\epsilon^{*}_{\varrho}(p_2)[2f_{B^{*}B^{*}V}(q^{\xi}g_{\varrho\kappa}-q^{\varrho}g_{\xi\kappa})\nonumber\\
             &+\frac{1}{2}g_{B^{*}B^{*}V}g^{\xi\varrho}(k_2+p_2)^{\kappa}]\frac{-g^{\beta\kappa}+q^{\beta}q^{\kappa}/m_V^2}{q^2-m_V^2},\\
{\cal{M}}_{b2}&=\frac{\sqrt{6}g_{TH}g_{\pi{}BB^{*}}g_{\Upsilon(11020)B_1\bar{B}^{*}}}{4\Lambda_{\chi}f_{\pi}}\int\frac{d^4q}{(2\pi)^4}\Phi[(k_1\omega_{\bar{B}^{*}}-k_2\omega_{B_1})_E^2]\nonumber\\
              &\times{}\epsilon_{\delta}^{*}(p_1)\gamma^{\delta}(1+v\!\!\!/)\gamma_5(g^{\mu\nu}-\frac{\gamma^{\nu}(\gamma^{\mu}-v^{\mu})}{3})q_{\mu}q\!\!\!/\gamma_5\epsilon_{\eta}(p)\epsilon_{\alpha\eta\beta\sigma}\nonumber\\
              &\times{}\frac{-g^{\nu\sigma}+k_1^{\nu}k_1^{\sigma}/m^2_{B_1}}{k_1^2-m^2_{B_1}}p^{\alpha}\frac{-g^{\lambda\beta}+k_2^{\lambda}k_2^{\beta}/m^2_{\bar{B}}}{k_2^2-m^2_{\bar{B}^{*}}}q^{\lambda}\frac{1}{q^2-m^2_{\pi}}\nonumber\\
              &+\sum_{V=\rho^0,\omega}i\frac{g_{B_1B^{*}V}g_{\Upsilon(11020)B_1\bar{B}}g_{BBV}}{2}\int\frac{d^4q}{(2\pi)^4}\Phi[(k_1\omega_{\bar{B}}-k_2\omega_{B_1})_E^2]\nonumber\\
              &\times{}\epsilon^{\mu\nu\alpha\beta}(p_1+k_1)_{\alpha}\epsilon_{\nu}^{*}(p_1)\frac{-g^{\mu\eta}+k_1^{\mu}k_1^{\eta}/m^2_{B_1}}{k_1^2-m^2_{B_1}}\epsilon_{\eta}(p)\frac{1}{k_2^2-m^2_{\bar{B}}}\nonumber\\
              &\times{}(k_2+p_2)_{\sigma}\frac{-g^{\beta\sigma}+q^{\beta}q^{\sigma}/m_V^2}{q^2-m_V^2}+\sum_{V=\rho^0,\omega}g_{B_1B^{*}V}g_{\Upsilon(11020)B_1\bar{B}^{*}}\nonumber\\
              &\times{}f_{B^{*}BV}\int\frac{d^4q}{(2\pi)^4}\Phi[(k_1\omega_{\bar{B}^{*}}-k_2\omega_{B_1})_E^2]\epsilon^{\mu\nu\alpha\beta}(p_1+k_1)_{\alpha}\nonumber\\
              &\times{}\epsilon_{\nu}^{*}(p_1)\frac{-g^{\mu\sigma}+k_1^{\mu}k_1^{\sigma}/m^2_{B_1}}{k_1^2-m^2_{B_1}}\epsilon_{\tau\eta\lambda\sigma}p^{\tau}\epsilon_{\eta}(p)\frac{-g^{\lambda\xi}+k_2^{\lambda}k_2^{\xi}/m^2_{\bar{B}^{*}}}{k_2^2-m^2_{\bar{B}^{*}}}\nonumber\\ &\times{}\epsilon_{\varphi\kappa\delta\xi}q^{\varphi}(p_2+k_2)^{\delta}\frac{-g^{\beta\kappa}+q^{\beta}q^{\kappa}/m_V^2}{q^2-m_V^2},\\
{\cal{M}}_{b3}&=\sum_{V=\rho^0,\omega}i\frac{g_{B_1BV}g_{VBB}g_{\Upsilon(11020)B_1\bar{B}}}{2}\int\frac{d^4q}{(2\pi)^4}\Phi[(k_1\omega_{\bar{B}}-k_2\omega_{B_1})_E^2]\nonumber\\
              &\times{}\frac{-g^{\mu\eta}+k_1^{\mu}k_1^{\eta}/m^2_{B_1}}{k_1^2-m^2_{B_1}}\epsilon_{\eta}(p)\frac{1}{k_2^2-m^2_{\bar{B}}}(k_2+p_2)_{\nu}\frac{-g^{\mu\nu}+q^{\mu}q^{\nu}/m_V^2}{q^2-m_V^2}\nonumber\\
              &-\sum_{V=\rho^0,\omega}g_{B_1BV}f_{VB^{*}B}g_{\Upsilon(11020)B_1\bar{B}^{*}}\int\frac{d^4q}{(2\pi)^4}\Phi[(k_1\omega_{\bar{B}^{*}}-k_2\omega_{B_1})_E^2]\nonumber\\
              &\times{}\frac{-g^{\mu\sigma}+k_1^{\mu}k_1^{\sigma}/m^2_{B_1}}{k_1^2-m^2_{B_1}}\epsilon_{\delta\eta\lambda\sigma}p^{\delta}\epsilon^{\eta}(p)\frac{-g^{\lambda\tau}+k_2^{\lambda}k_2^{\tau}/m^2_{\bar{B}^{*}}}{k_2^2-m^2_{\bar{B}^{*}}}\epsilon_{\nu\alpha\beta\tau}q^{\nu}\nonumber\\
              &\times{}(p_2+k_2)^{\beta}\frac{-g^{\mu\alpha}+q^{\mu}q^{\alpha}/m_V^2}{q^2-m_V^2},\\
{\cal{M}}_{b4}&=\frac{-i\sqrt{6}g_{TH}g_{BB^{*}M}g_{\Upsilon(11020)B_1\bar{B}}}{\Lambda_{\chi}f_{\pi}}\int\frac{d^4q}{(2\pi)^4}\Phi[(k_1\omega_{\bar{B}}-k_2\omega_{B_1})_E^2]\nonumber
\end{align}
\begin{align}
                &\times{}\epsilon_{\alpha}^{*}(p_1)\gamma^{\alpha}(1+v\!\!\!/)\gamma_5(g^{\mu\nu}-\frac{\gamma^{\nu}(\gamma^{\mu}-v^{\mu})}{3})q_{\mu}q\!\!\!/\gamma_5\epsilon_{\eta}(p)\nonumber\\
               &\times\frac{-g^{\nu\eta}+k_1^{\nu}k_1^{\eta}/m^2_{B_1}}{k_1^2-m^2_{B_1}}\frac{1}{k_2^2-m^2_{\bar{B}}}\epsilon^{*}_{\sigma}(p_2)q^{\sigma}\frac{1}{q^2-m^2_{K}}-i2g_{B_1B^{*}V}\nonumber\\
               &\times{}f_{B^{*}BV}g_{\Upsilon(11020)B_1\bar{B}}\int\frac{d^4q}{(2\pi)^4}\Phi[(k_1\omega_{\bar{B}}-k_2\omega_{B_1})_E^2]\epsilon_{\mu\nu\alpha\beta}\nonumber\\
               &\times{}(p_1+k_1)^{\alpha}\epsilon^{*\nu}(p_1)\frac{-g^{\mu\eta}+k_1^{\mu}k_1^{\eta}/m^2_{B_1}}{k_1^2-m^2_{B_1}}\epsilon_{\eta}(p)\frac{1}{k_2^2-m^2_{\bar{B}}}\nonumber\\
               &\times{}\epsilon_{\sigma\lambda\delta\tau}\epsilon^{*\tau}(p_2)(k_2+p_2)^{\delta}\frac{-g^{\beta\lambda}+q^{\beta}q^{\lambda}/m^2_{K^{*}}}{q^2-m^2_{K^{*}}}q^{\sigma}+\sqrt{6}g_{TH}\nonumber\\
               &\times\frac{g_{B^{*}B^{*}M}g_{\Upsilon(11020)B_1\bar{B}^{*}}}{2\Lambda_{\chi}f_{\pi}}\int\frac{d^4q}{(2\pi)^4}\Phi[(k_1\omega_{\bar{B}^{*}}-k_2\omega_{B_1})_E^2]\nonumber\\
               &\times{}\epsilon_{\alpha}^{*}(p_1)\gamma^{\alpha}(1+v\!\!\!/)\gamma_5(g^{\mu\nu}-\frac{\gamma^{\nu}(\gamma^{\mu}-v^{\mu})}{3})q_{\mu}q\!\!\!/\gamma_5\epsilon_{\eta}(p)\nonumber\\
               &\times\frac{-g^{\nu\sigma}+k_1^{\nu}k_1^{\sigma}/m^2_{B_1}}{k_1^2-m^2_{B_1}}\epsilon_{\lambda\eta\beta\sigma}p^{\lambda}\frac{-g^{\beta\tau}+k_2^{\beta}k_2^{\tau}/m^2_{\bar{B}^{*}}}{k_2^2-m^2_{\bar{B}^{*}}}\epsilon_{\xi\delta\varphi\tau}\nonumber\\
               &\times{}q^{\delta}(p_2+q)^{\varphi}\frac{1}{q^2-m_K^2}\epsilon^{*\xi}(p_2)-g_{B_1B^{*}V}g_{\Upsilon(11020)B_1\bar{B}^{*}}\nonumber\\
               &\times\int\frac{d^4q}{(2\pi)^4}\Phi[(k_1\omega_{\bar{B}^{*}}-k_2\omega_{B_1})_E^2]\epsilon_{\mu\nu\alpha\beta}(p_1+k_1)^{\alpha}\epsilon^{*\nu}(p_1)\nonumber\\
               &\times{}\frac{-g^{\mu\sigma}+k_1^{\mu}k_1^{\sigma}/m^2_{B_1}}{k_1^2-m^2_{B_1}}\epsilon_{\eta}(p)\epsilon_{\lambda\eta\delta\sigma}p^{\lambda}\frac{-g^{\delta\tau}+k_2^{\delta}k_2^{\tau}/m^2_{\bar{B}^{*}}}{k_2^2-m^2_{\bar{B}^{*}}}\nonumber\\
               &\times{}\frac{-g^{\beta\pi}+q^{\beta}q^{\pi}/m^2_{K^{*}}}{q^2-m^2_{K^{*}}}\epsilon_{\varphi}^{*}(p_2)[g_{B^{*}B^{*}V}g_{\tau\varphi}(p_2+k_2)_{\pi}\nonumber\\
               &-4f_{B^{*}B^{*}V}(q_{\tau}g_{\varphi\pi}-q_{\varphi}g_{\tau\pi})],\\
{\cal{M}}_{b5}&=ig_{B_1B^{*}V}g_{BBV}g_{\Upsilon(11020)B_1\bar{B}}\int\frac{d^4q}{(2\pi)^4}\Phi[(k_1\omega_{\bar{B}}-k_2\omega_{B_1})_E^2]\nonumber\\
              &\times{}\epsilon_{\mu\nu\alpha\beta}(p_1+k_1)^{\alpha}\epsilon^{*\nu}(p_1)\frac{-g^{\mu\eta}+k_1^{\mu}k_1^{\eta}/m^2_{B_1}}{k_1^2-m^2_{B_1}}\epsilon_{\eta}(p)\frac{1}{k_2^2-m^2_{\bar{B}}}\nonumber\\
              &\times{}(k_2+p_2)^{\lambda}\frac{-g^{\beta\lambda}+q^{\beta}q^{\lambda}/m^2_{K^{*}}}{q^2-m^2_{K^{*}}}-\frac{\sqrt{6}g_{TH}g_{B^{*}BM}g_{\Upsilon(11020)B_1\bar{B}^{*}}}{2\Lambda_{\chi}f_{\pi}}\nonumber\\
              &\times\int\frac{d^4q}{(2\pi)^4}\Phi[(k_1\omega_{\bar{B}^{*}}-k_2\omega_{B_1})_E^2]\epsilon_{\alpha}^{*}(p_1)\gamma^{\alpha}(1+v\!\!\!/)\gamma_5(g^{\mu\nu}\nonumber\\
              &-\frac{\gamma^{\nu}(\gamma^{\mu}-v^{\mu})}{3})q_{\mu}q\!\!\!/\gamma_5\epsilon_{\eta}(p)\frac{-g^{\nu\sigma}+k_1^{\nu}k_1^{\sigma}/m^2_{B_1}}{k_1^2-m^2_{B_1}}\epsilon_{\lambda\eta\beta\sigma}p^{\lambda}\nonumber\\
              &\times\frac{-g^{\beta\tau}+k_2^{\beta}k_2^{\tau}/m^2_{\bar{B}^{*}}}{k_2^2-m^2_{\bar{B}^{*}}}q_{\tau}\frac{1}{q^2-m^2_K}+2g_{B_1B^{*}V}f_{B^{*}BV}g_{\Upsilon(11020)B_1\bar{B}^{*}}\nonumber\\
              &\times\int\frac{d^4q}{(2\pi)^4}\Phi[(k_1\omega_{\bar{B}^{*}}-k_2\omega_{B_1})_E^2]\epsilon_{\mu\nu\alpha\beta}(p_1+k_1)^{\alpha}\epsilon^{*\nu}(p_1)\nonumber\\
              &\times{}\frac{-g^{\mu\sigma}+k_1^{\mu}k_1^{\sigma}/m^2_{B_1}}{k_1^2-m^2_{B_1}}\epsilon_{\eta}(p)\epsilon_{\lambda\eta\delta\sigma}p_{\lambda}\frac{-g^{\delta\tau}+k_2^{\delta}k_2^{\tau}/m^2_{\bar{B}^{*}}}{k_2^2-m^2_{\bar{B}^{*}}}\nonumber\\
              &\times{}\epsilon_{\xi\varphi\pi\tau}(k_2+p_2)^{\pi}q^{\xi}\frac{-g^{\beta\varphi}+q^{\beta}q^{\varphi}/m^2_{K^{*}}}{q^2-m^2_{K^{*}}},\\
{\cal{M}}_{b6}&=-ig_{B_1BV}g_{BBV}g_{\Upsilon(11020)B_1\bar{B}}\int\frac{d^4q}{(2\pi)^4}\Phi[(k_1\omega_{\bar{B}}-k_2\omega_{B_1})_E^2]\nonumber
\end{align}
\begin{align}
              &\times\frac{-g^{\mu\eta}+k_1^{\mu}k_1^{\eta}/m^2_{B_1}}{k_1^2-m^2_{B_1}}\epsilon_{\eta}(p)\frac{1}{k_2^2-m^2_{\bar{B}}}(p_2+k_2)_{\nu}\frac{-g^{\mu\nu}+q^{\mu}q^{\nu}/m^2_{K^{*}}}{q^2-m^2_{K^{*}}}\nonumber\\
              &-2g_{B_1BV}f_{B^{*}BV}g_{\Upsilon(11020)B_1\bar{B}^{*}}\int\frac{d^4q}{(2\pi)^4}\Phi[(k_1\omega_{\bar{B}^{*}}-k_2\omega_{B_1})_E^2]\nonumber\\
              &\times\frac{-g^{\mu\sigma}+k_1^{\mu}k_1^{\sigma}/m^2_{B_1}}{k_1^2-m^2_{B_1}}\epsilon_{\nu\eta\alpha\sigma}p^{\nu}\frac{-g^{\alpha\lambda}+k_2^{\alpha}k_2^{\lambda}/m^2_{\bar{B}^{*}}}{k_2^2-m^2_{\bar{B}^{*}}}\epsilon_{\delta\xi\varphi\lambda}q^{\delta}\nonumber\\
              &\times(k_2+p_2)^{\varphi}\frac{-g^{\mu\xi}+q^{\mu}q^{\xi}/m^2_{K^{*}}}{q^2-m^2_{K^{*}}},\\
{\cal{M}}_{c}&=-\frac{\sqrt{3}g_{\Upsilon(11020)B_1\bar{B}}g_{Z_b\Upsilon(nS)\pi}g_{TH}g_{Z_bBB^{*}}}{4\Lambda_{\chi}f_{\pi}m_{Z_b}}\int\frac{d^4q}{(2\pi)^4}\nonumber\\
             &\times\Phi[(k_1\omega_{\bar{B}}-k_2\omega_{B_1})_E^2][p_2\cdot{}p_3\epsilon^{*}_{\alpha}(p_3)-p_2\cdot\epsilon^{*}(p_3)\nonumber\\
             &\times{}p_{3\alpha}]\frac{-g^{\alpha\beta}+p_l^{\alpha}p_l^{\beta}/m^2_{Z_b}}{p_l^2-m^2_{Z_b}+im_{Z_b}\Gamma_{Z_b}}\frac{-g^{\beta\eta}+q^{\beta}q^{\eta}/m^2_{B^{*}}}{q^2-m^2_{B^{*}}}\gamma_{\eta}(1+v\!\!\!/)\nonumber\\
             &\times\gamma_5(g^{\lambda\sigma}-\frac{\gamma^{\sigma}(\gamma^{\lambda}-v^{\lambda})}{3})p_{1\lambda}p\!\!\!/_1\gamma_5\frac{-g^{\sigma\tau}+k_1^{\sigma}k_1^{\tau}/m^2_{B_1}}{k_1^2-m^2_{B_1}}\nonumber\\
             &\times\frac{1}{k_2^2-m^2_{\bar{B}}}\epsilon_{\tau}(p)-\frac{\sqrt{3}g_{\Upsilon(11020)B_1\bar{B}^{*}}g_{Z_b^{'}\Upsilon(nS)\pi}g_{TH}g_{Z_b^{'}B^{*}\bar{B}^{*}}}{4\Lambda_{\chi}f_{\pi}m_{Z_b^{'}}}\nonumber\\
             &\times\int\frac{d^4q}{(2\pi)^4}\Phi[(k_1\omega_{\bar{B}^{*}}-k_2\omega_{B_1})_E^2][p_2\cdot{}p_3\epsilon^{*}_{\delta}(p_3)-p_2\cdot\epsilon^{*}(p_3)\nonumber\\
             &\times{}p_{\delta}]\frac{-g^{\delta\sigma}+p_l^{\delta}p_l^{\sigma}/m^2_{Z_b^{'}}}{p_l^2-m^2_{Z_b}+im_{Z_b^{'}}\Gamma_{Z_b^{'}}}\epsilon_{\tau\sigma\alpha\beta}p_{l}^{\tau}\frac{-g^{\alpha\varphi}+q^{\alpha}q^{\varphi}/m^2_{B^{*}}}{q^2-m^2_{B^{*}}}\gamma_{\varphi}\nonumber\\
             &\times(1+v\!\!\!/)\gamma_5(g^{\xi\phi}-\frac{\gamma^{\phi}(\gamma^{\xi}-v^{\xi})}{3})p_{1\xi}p\!\!\!/_1\gamma_5\frac{-g^{\phi\pi}+k_1^{\phi}k_1^{\pi}/m^2_{B_1}}{k_1^2-m^2_{B_1}}\nonumber\\
             &\times{}\epsilon_{\lambda\eta\theta\pi}p^{\lambda}\epsilon^{\eta}(p)\frac{-g^{\theta\beta}+k_2^{\theta}k_2^{\beta}/m^2_{\bar{B}^{*}}}{k_2^2-m^2_{\bar{B}^{*}}},\\
{\cal{M}}_{d}&=-\frac{\sqrt{3}g_{\Upsilon(11020)B_1\bar{B}}g_{Z_b(10610)\Upsilon(nS)\pi}g_{TH}g_{Z_bBB^{*}}}{4\Lambda_{\chi}f_{\pi}m_{Z_b}}\int\frac{d^4q}{(2\pi)^4}\nonumber\\
             &\times\Phi[(k_1\omega_{\bar{B}}-k_2\omega_{B_1})_E^2]\epsilon^{\mu\nu\alpha\beta}p_{3\nu}\epsilon^{*}_{\alpha}(p_3)p_{2\beta}\nonumber\\
             &\times{}\frac{-g^{\mu\sigma}+p_l^{\mu}p_l^{\sigma}/m^2_{Z_b}}{p_l^2-m^2_{Z_b}+im_{Z_b}\Gamma_{Z_b}}\frac{-g^{\sigma\eta}+q^{\sigma}q^{\eta}/m^2_{B^{*}}}{q^2-m^2_{B^{*}}}\gamma_{\eta}(1+v\!\!\!/)\gamma_5\nonumber\\
             &\times(g^{\lambda\tau}-\frac{\gamma^{\tau}(\gamma^{\lambda}-v^{\lambda})}{3})p_{1\lambda}p\!\!\!/_1\gamma_5\frac{-g^{\tau\varpi}+k_1^{\tau}k_1^{\varpi}/m^2_{B_1}}{k_1^2-m^2_{B_1}}\nonumber\\
             &\times\frac{1}{k_2^2-m^2_{\bar{B}}}\epsilon_{\varpi}(p)-\frac{\sqrt{3}g_{\Upsilon(11020)B_1\bar{B}^{*}}g_{Z_b^{'}\Upsilon(nS)\pi}g_{TH}g_{Z_b^{'}BB^{*}}}{4\Lambda_{\chi}f_{\pi}m_{Z_b^{'}}}\nonumber\\
             &\times\int\frac{d^4q}{(2\pi)^4}\Phi[(k_1\omega_{\bar{B}^{*}}-k_2\omega_{B_1})_E^2]\epsilon^{\rho\varrho\varpi\kappa}p_{3\varrho}\epsilon^{*}_{\varpi}(p_3)p_{2\kappa}\nonumber\\
             &\times{}\frac{-g^{\rho\sigma}+p_l^{\rho}p_l^{\sigma}/m^2_{Z_b^{'}}}{p_l^2-m^2_{Z_b^{'}}+im_{Z_b^{'}}\Gamma_{Z_b^{'}}}\epsilon_{\tau\sigma\alpha\beta}p_{l}^{\tau}\frac{-g^{\alpha\varphi}+q^{\alpha}q^{\varphi}/m^2_{B^{*}}}{q^2-m^2_{B^{*}}}\nonumber\\
             &\times\gamma_{\varphi}(1+v\!\!\!/)\gamma_5(g^{\xi\phi}-\frac{\gamma^{\phi}(\gamma^{\xi}-v^{\xi})}{3})p_{1\xi}p\!\!\!/_1\gamma_5\frac{-g^{\phi\pi}+k_1^{\phi}k_1^{\pi}/m^2_{B_1}}{k_1^2-m^2_{B_1}}\nonumber\\
             &\times{}\epsilon_{\lambda\eta\theta\pi}p^{\lambda}\epsilon^{\eta}(p)\frac{-g^{\theta\beta}+k_2^{\theta}k_2^{\beta}/m^2_{\bar{B}^{*}}}{k_2^2-m^2_{\bar{B}^{*}}},
\end{align}
\begin{align}
{\cal{M}}_{e1}&=i\frac{\sqrt{3}g_{TH}g_{\Upsilon(11020)B_1\bar{B}}}{4\Lambda_{\chi}f_{\pi}}\Phi[(k_1\omega_{\bar{B}}-k_2\omega_{B_1})_E^2]\epsilon_{\mu}(p)\nonumber\\
             &\times\frac{-g_{\mu\nu}+k_{1\mu}k_{1\mu}/m^2_{B_1}}{k_1^2-m^2_{B_1}}\epsilon^{*}_{\sigma}(p_1)\gamma^{\sigma}(1+v\!\!\!/)\gamma_5\nonumber\\
             &\times(g^{\alpha\nu}-\frac{\gamma^{\nu}(\gamma^{\alpha}-v^{\alpha})}{3})p_{2\alpha}p\!\!\!/_2\gamma_5,\\
{\cal{M}}_{e2}&=-i\frac{\sqrt{3}g_{TH}g_{\Upsilon(11020)B_1\bar{B}^{*}}}{4\Lambda_{\chi}f_{\pi}}\Phi[(k_1\omega_{\bar{B}^{*}}-k_2\omega_{B_1})_E^2]\nonumber\\
             &\times{}\epsilon^{*}_{\sigma}(p_1)\gamma^{\sigma}(1+v\!\!\!/)\gamma_5(g^{\mu\nu}-\frac{\gamma^{\nu}(\gamma^{\mu}-v^{\mu})}{3})p_{2\mu}p\!\!\!/_2\gamma_5\nonumber\\
             &\times{}\frac{-g^{\nu\sigma}+k_1^{\nu}k_1^{\sigma}/m^2_{B_1}}{k_1^2-m^2_{B_1}}\epsilon_{\alpha\eta\beta\sigma}p^{\alpha}\epsilon^{\eta}(p)\epsilon^{*\beta}(k_2),\\
{\cal{M}}_{f1}&=i\frac{g_{\chi_{b0}B\bar{B}}g_{B_1BV}g_{\Upsilon(11020)B_1\bar{B}}}{\sqrt{2}}\int\frac{d^4q}{(2\pi)^4}\Phi[(k_1\omega_{\bar{B}}-k_2\omega_{B_1})_E^2]\nonumber\\
              &\times\epsilon^{*}_{\mu}(p_t)\frac{-g^{\mu\nu}+k_1^{\mu}k_1^{\nu}/m^2_{B_1}}{k_1^2-m^2_{B_1}}\epsilon_{\nu}(p)\frac{1}{k_2^2-m^2_{\bar{B}}}\frac{1}{q^2-m_B^2}\nonumber\\
              &-\frac{g_{\chi_{b0}B^{*}\bar{B}^{*}}g_{B_1B^{*}V}g_{\Upsilon(11020)B_1\bar{B}^{*}}}{\sqrt{2}}\int\frac{d^4q}{(2\pi)^4}\Phi[(k_1\omega_{\bar{B}^{*}}-k_2\omega_{B_1})_E^2]\nonumber\\
              &\times{}\epsilon_{\mu\nu\alpha\beta}(k_1+q)^{\alpha}\epsilon^{*\beta}(p_t)\frac{-g^{\mu\sigma}+k_1^{\mu}k_1^{\sigma}/m^2_{B_1}}{k_1^2-m^2_{B_1}}\epsilon_{\tau\eta\lambda\sigma}p^{\tau}\epsilon^{\eta}(p)\nonumber\\
                           &\times\frac{-g^{\lambda\xi}+k_2^{\lambda}k_2^{\xi}/m^2_{\bar{B}^{*}}}{k_2^2-m^2_{\bar{B}^{*}}}\frac{-g^{\nu\xi}+q^{\nu}q^{\xi}/m^2_{B^{*}}}{q^2-m^2_{B^{*}}}\label{eq24},\\
{\cal{M}}_{f2}&=-\frac{g_{\chi_{b1}\bar{B}B^{*}}g_{B_1B^{*}V}g_{\Upsilon(11020)B_1\bar{B}}}{\sqrt{2}}\int\frac{d^4q}{(2\pi)^4}\Phi[(k_1\omega_{\bar{B}}-k_2\omega_{B_1})_E^2]\nonumber\\
               &\times\epsilon_{\mu\nu\alpha\beta}(q+k_1)^{\alpha}\epsilon^{*\beta}(p_t)\epsilon_{\eta}(p)\epsilon^{*}_{\lambda}(p_4)\frac{-g^{\lambda\nu}+q^{\lambda}q^{\nu}/m^2_{B^{*}}}{q^2-m^2_{B^{*}}}\nonumber\\
             &\times\frac{-g^{\mu\eta}+k_1^{\mu}k_1^{\eta}/m^2_{B_1}}{k_1^2-m^2_{B_1}}\frac{1}{k_2^2-m^2_{\bar{B}}}-i\frac{g_{\chi_{b1}\bar{B}B^{*}}g_{B_1BV}g_{\Upsilon(11020)B_1\bar{B}^{*}}}{\sqrt{2}}\nonumber\\
             &\int\frac{d^4q}{(2\pi)^4}\Phi[(k_1\omega_{\bar{B}^{*}}-k_2\omega_{B_1})_E^2]\epsilon^{*}_{\mu}(p_t)\frac{-g^{\mu\sigma}+k_1^{\mu}k_1^{\sigma}/m^2_{B_1}}{k_1^2-m^2_{B_1}}\nonumber\\
             &\times{}\epsilon_{\lambda\eta\alpha\sigma}p^{\lambda}\epsilon^{\eta}(p)\frac{-g^{\alpha\delta}+k_2^{\alpha}k_2^{\delta}/m^2_{\bar{B}^{*}}}{k_2^2-m^2_{\bar{B}^{*}}}\epsilon^{*}_{\delta}(p_4)\frac{1}{q^2-m^2_{B}},\\
{\cal{M}}_{f3}&=-\frac{ig_{\chi_{b2}BB}g_{B_1BV}g_{\Upsilon(11020)B_1\bar{B}}}{\sqrt{2}}\int\frac{d^4q}{(2\pi)^4}\Phi[(k_1\omega_{\bar{B}}-k_2\omega_{B_1})_E^2]\nonumber\\
             &\times{}\epsilon_{\beta}(p)\epsilon^{*}_{\mu\nu}(p_4)q^{\mu}q^{\nu}\frac{1}{q^2-m_B^2}\epsilon^{*\alpha}(p_t)\frac{-g^{\alpha\beta}+k_1^{\alpha}k_1^{\beta}/m^2_{B_1}}{k_1^2-m^2_{B_1}}\nonumber\\
              &\times\frac{1}{k_2^2-m^2_{\bar{B}}}-\frac{ig_{\chi_{b2}B^{*}B}g_{B_1B^{*}V}g_{\Upsilon(11020)B_1\bar{B}}}{\sqrt{2}}\int\frac{d^4q}{(2\pi)^4}\nonumber\\
              &\times\Phi[(k_1\omega_{\bar{B}}-k_2\omega_{B_1})_E^2]\epsilon_{\xi}(p)\epsilon^{*\tau}(p_t)\epsilon_{\mu\nu\alpha\beta}p^{\alpha}_{4}q^{\rho}k_{2\beta}\epsilon_{\mu\rho}(p_4)\nonumber\\
              &\times{}\frac{-g^{\nu\eta}+q^{\nu}q^{\eta}/m^2_{B^{*}}}{q^2-m^2_{B^{*}}}(q+k_{1})^{\lambda}\epsilon_{\sigma\eta\lambda\tau}\frac{-g^{\sigma\xi}+k_1^{\sigma}k_1^{\xi}/m^2_{B_1}}{k_1^2-m^2_{B_1}}\nonumber\\
             &\times\frac{1}{k_2^2-m^2_{\bar{B}}}+\frac{g_{\chi_{b2}B^{*}B}g_{B_1BV}g_{\Upsilon(11020)B_1\bar{B}^{*}}}{\sqrt{2}}\int\frac{d^4q}{(2\pi)^4}\nonumber\\
             &\times\Phi[(k_1\omega_{\bar{B}^{*}}-k_2\omega_{B_1})_E^2]\epsilon^{*}_{\mu}(p_t)\frac{-g^{\mu\sigma}+k_1^{\mu}k_1^{\sigma}/m^2_{B_1}}{k_1^2-m^2_{B_1}}\epsilon_{\tau\eta\lambda\sigma}\nonumber
\end{align}
\begin{align}
             &\times{}p^{\tau}\epsilon^{\eta}(p)\frac{-g^{\lambda\xi}+k_2^{\lambda}k_2^{\xi}/m^2_{\bar{B}^{*}}}{k_2^2-m^2_{\bar{B}^{*}}}\epsilon_{\pi\xi\delta\varphi}p_4^{\delta}\epsilon^{*\pi\chi}(p_4)k_{2\chi}q^{\varphi}\frac{1}{q^2-m^2_{B}}\nonumber\\
             &-\frac{g_{\chi_{b2}B^{*}B^{*}}g_{B_1B^{*}V}g_{\Upsilon(11020)B_1\bar{B}^{*}}}{\sqrt{2}}\int\frac{d^4q}{(2\pi)^4}\Phi[(k_1\omega_{\bar{B}^{*}}-k_2\omega_{B_1})_E^2]\nonumber\\
             &\times{}\epsilon_{\mu\nu\alpha\beta}\epsilon^{*\beta}(p_t)(q+k_1)^{\alpha}\frac{-g^{\mu\sigma}+k_1^{\mu}k_1^{\sigma}/m^2_{B_1}}{k_1^2-m^2_{B_1}}\epsilon_{\tau\eta\lambda\sigma}p^{\tau}\epsilon^{\eta}(p)\nonumber\\
             &\times\frac{-g^{\lambda\xi}+k_2^{\lambda}k_2^{\xi}/m^2_{\bar{B}^{*}}}{k_2^2-m^2_{\bar{B}^{*}}}\epsilon^{*\pi\xi}(p_4)\frac{-g^{\pi\nu}+q^{\pi}q^{\nu}/m^2_{B^{*}}}{q^2-m^2_{B^{*}}}\label{eq26},
\end{align}
where \( p_t = p_1 + p_2 + p_3 \) and \( p_l = p_2 + p_3 \).  The symbol \( \epsilon(p) \) denotes the polarization vector of the initial \(\Upsilon(11020)\) state, while \( \epsilon^{*} \) corresponds to that of the final states. The amplitudes \({\cal M}_{b1}\), \({\cal M}_{b2}\), and \({\cal M}_{b3}\) represent the transitions to the \( B^{*}\bar{B}^{*} \), \( B^{*}\bar{B} \), and \( B\bar{B} \) final states, respectively.
Similarly, \({\cal M}_{b4}\), \({\cal M}_{b5}\), and \({\cal M}_{b6}\) correspond to the \( B_s^{*}\bar{B}_s^{*} \), \( B_s^{*}\bar{B}_s \), and \( B_s\bar{B}_s \) final states, respectively.
The amplitudes \({\cal M}_{e1}\) and \({\cal M}_{e2}\) describe the three-body final states \( B^{*}\pi\bar{B} \) and \( B^{*}\pi\bar{B}^{*} \), respectively.
In addition, \({\cal M}_{f1}\), \({\cal M}_{f2}\), and \({\cal M}_{f3}\) denote the amplitudes for the final states involving the production of \( \chi_{b0} \), \( \chi_{b1} \), and \( \chi_{b2} \), respectively.  For the production of \( Z_b^{(\prime)} \), only the \( B_1^0\bar{B}^{(*)0} \) component contributes.  Therefore, the total amplitude should be multiplied by a factor of \(1/\sqrt{2}\), which originates from the \(C\)-parity of the \(\Upsilon(11020)\) and leads to its flavor wave function being expressed as~\cite{Yue:2024bvy}
\begin{align}
|\Upsilon(11020)\rangle = 1/\sqrt{2}\left(|B_1^{+}B^{(*)-}\rangle + |B_1^{0}\bar{B}^{(*)0}\rangle\right).
\end{align}

Once the amplitudes are obtained, they can be substituted into decay width formulas to calculate the corresponding partial widths. The detailed expressions can be found in Ref.~\cite{ParticleDataGroup:2024cfk}.  For both two-body and three-body decay processes, the final expressions are obtained by performing direct integrations over the respective
phase spaces, as given below
\begin{align}
d\Gamma(\Upsilon(11020)\to{}AB)&=\frac{1}{2J+1}\frac{1}{32\pi^2}\frac{|\vec{p}_A|}{m^2_{\Upsilon(11020)}}|\bar{{\cal{M}}}|^2d\Omega,\\
d\Gamma(\Upsilon(11020)\to{}AB&\to{}ACD)=\frac{1}{2J+1}\frac{1}{(2\pi)^5}\frac{1}{16m^2_{\Upsilon(11020)}}\nonumber\\
                              &\times|\bar{{\cal{M}}}|^2|\vec{p}^{*}_C||\vec{p}_A|dm_{CD}d\Omega^{*}_{p_C}d\Omega_{p_A},
\end{align}
where $J$ denotes the total angular momentum of $\Upsilon(11020)$, $|\vec{p}_A|$ represents the three-momentum of the decay products in the center-of-mass frame,
and the overline indicates a sum over the polarization vectors of the final hadrons. The quantities $(\vec{p}_C^*, \Omega_C^*)$ correspond to the momentum and emission
solid angle of the particle $C$ in the rest frame of the $CD$ system, while $\Omega_{p_A}$ denotes the solid angle of particle $A$ in the rest frame of the decaying particle.
The variable $m_{CD}$ is the invariant mass of the $CD$ pair, subject to the constraint $ m_C + m_D \le m_{CD} \le m_{\Upsilon(11020)} - m_{A}$.

Using the two-body decay formulas obtained above, the decay width for $\Upsilon(11020)\to \chi_{bJ}\omega \to \chi_{bJ}\pi\pi\pi$ can be factorized as
\begin{align}
\Gamma(\Upsilon(11020) &\to \chi_{bJ}\omega \to \chi_{bJ}\pi\pi\pi)\nonumber\\
&= \Gamma(\Upsilon(11020)\to \chi_{bJ}\omega)
\times {\rm Br}(\omega\to 3\pi),
\end{align}
where ${\rm Br}(\omega\to 3\pi)=(89.2\pm 0.7)\%$~\cite{ParticleDataGroup:2024cfk} denotes the branching ratio of the $\omega$ meson decaying into three pions.  A detailed calculation
is provided in Appendix.~\ref{app:decay_width}, along with a similar example for which the cross-section formula for the corresponding multi-body final state can be found in
Ref.~\cite{Nam:2015yoa}.   $\Gamma(\Upsilon(11020)\to \chi_{bJ}\omega)$ denotes the two-body decay width of $\Upsilon(11020)\to \chi_{bJ}\omega$,
whose corresponding amplitudes have been given in Eqs.~(\ref{eq24}--\ref{eq26}).

\section{RESULTS AND DISCUSSIONS}\label{Sec: results}
Prior to evaluating the decay widths, the unknown coupling constants \( g_{\Upsilon(11020)B_1B^{(*)}} \) and \( g_{B_1B^{*}\gamma} \) associated with the effective Lagrangians, as well as the
model parameter \( \Lambda \), need to be determined.  To this end, we perform a fit to the experimental partial widths of the decays \(\Upsilon(11020) \to e^{+}e^{-}\) and \(\Upsilon(11020)
\to \pi\pi\pi\chi_{bJ}(1P)\), which correspond to branching fractions of \((5.4^{+1.9}_{-2.1}) \times 10^{-6}\) and \(9^{+9}_{-8}\times 10^{-3}\), respectively~\cite{ParticleDataGroup:2024cfk}.
It should be noted that the decay width of \(\Upsilon(11020) \to \pi\pi\pi\chi_{bJ}(1P)\) is regarded as the total of the widths for the three channels \(\chi_{b0}(1P)\),
\(\chi_{b1}(1P)\), and \(\chi_{b2}(1P)\).
\begin{figure}[h!]
\begin{center}
\includegraphics[bb=20 126 650 370, clip, scale=0.57]{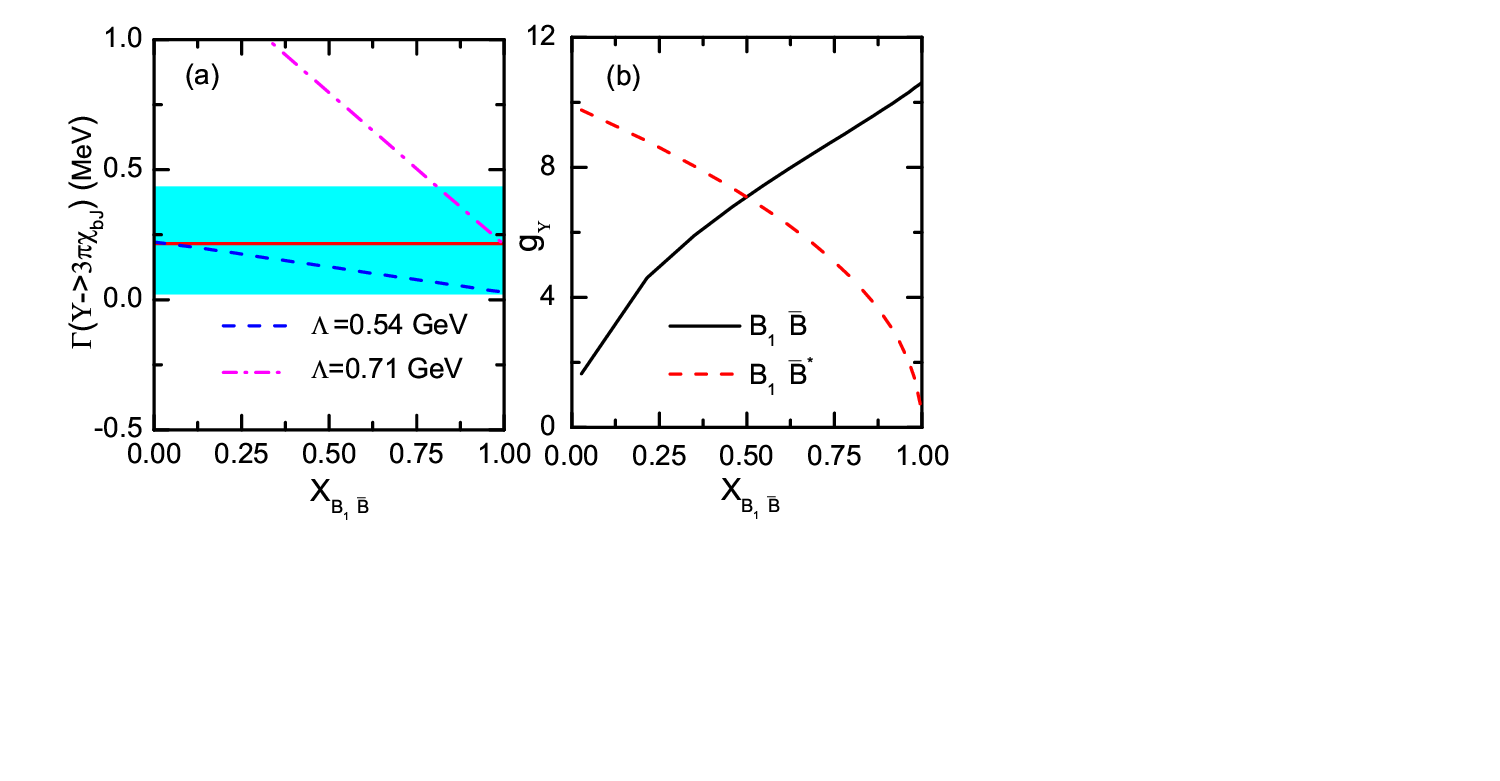}
\caption{(a) Decay width of $\Upsilon(11020)\to \pi\pi\pi\chi_{bJ}(1P)$ as a function of the model parameter $\Lambda$ and the $B_1\bar{B}$ molecular component. The cyan band indicates the experimental uncertainty, while the red solid line denotes the central value.  (b) Coupling constants $g_Y$ (in GeV for $\Upsilon(11020)$--$B_1\bar{B}$ and in GeV$^{-1}$ for $\Upsilon(11020)$--$B_1\bar{B}^{*}$) as a function of the molecular component, evaluated at the values of $\Lambda$ corresponding to the molecular components listed in Table~\ref{table2}.}
\label{cc3}
\end{center}
\end{figure}

Within the molecular scenario, where the $\Upsilon(11020)$ is interpreted as a bound state of $B_1\bar{B}$ and $B_1\bar{B}^*$, the decay amplitudes entering the fit are given in Eqs.~(\ref{eq23}--\ref{eq26}). Using these amplitudes, we obtain the $\Upsilon(11020)\to \pi\pi\pi\chi_{bJ}(1P)$ decay width shown in Fig.~\ref{cc3}(a), which illustrates its dependence on the molecular component $X_{B_1\bar{B}}$ and the cutoff parameter $\Lambda$. As seen in the figure, when only the experimental central value of the decay width (0.216 MeV) is employed, $X_{B_1\bar{B}}$ can vary over the full range from 0 to 1, while $\Lambda$ is restricted to the interval 0.54--0.71~GeV.  Table~\ref{table2} summarizes the fitted values of $X_{B_1\bar{B}}$ as a function of $\Lambda$ (in steps of 0.01), evaluated at a fixed width of 0.216~MeV for the $\Upsilon(11020)\to \pi\pi\pi\chi_{bJ}(1P)$ decay. We note that the cyan band in Fig.~\ref{cc3} (a) dot included in the fit. Incorporating the full experimental uncertainties from both the total width and the branching fraction would substantially enlarge the propagated errors in the extracted parameters, and is therefore avoided in the present analysis.
\begin{table}[h!]
\centering
\tabcolsep=3.5mm
\caption{Fitted values of the model parameters and the $B_{1}\bar{B}$ molecular component for a $\Upsilon(11020)\to \pi\pi\pi\chi_{bJ}(1P)$ decay width of 0.216~MeV.}
\label{table2}
\begin{tabular}{cc|cc|cc}
\hline\hline
$\Lambda$ & $X_{B_1\bar{B}}$ & $\Lambda$ & $X_{B_1\bar{B}}$ & $\Lambda$ & $X_{B_1\bar{B}}$ \\
\hline
0.54 & 0.0279 & 0.60 & 0.684 & 0.66 & 0.915 \\
0.55 & 0.215  & 0.61 & 0.737 & 0.67 & 0.940 \\
0.56 & 0.350  & 0.62 & 0.781 & 0.68 & 0.963 \\
0.57 & 0.459  & 0.63 & 0.818 & 0.69 & 0.977 \\
0.58 & 0.548  & 0.64 & 0.856 & 0.70 & 0.996 \\
0.59 & 0.622  & 0.65 & 0.887 & 0.71 & 1.017 \\
\hline\hline
\end{tabular}
\end{table}

Using the fitted values of the model parameter $\Lambda$ and the corresponding molecular components listed in Table~\ref{table2}, we calculate the corresponding coupling constants \( g_{\Upsilon(11020)B_1B^{(*)}} \). The resulting values are shown in Fig.~\ref{cc3} (b).  We find that the coupling \( g_{\Upsilon(11020)B_1\bar{B}} \) increases monotonically
with increasing $X_{B_1\bar{B}}$. Comparing \( g_{\Upsilon(11020)B_1\bar{B}} \) with \( g_{\Upsilon(11020)B_1\bar{B}^{*}} \), we find that their behaviors are quite different: the coupling constant \( g_{\Upsilon(11020)B_1\bar{B}^{*}} \) decreases with increasing $X_{B_1\bar{B}}$.  The opposite trend can be easily understood, as the coupling constants \( g_{\Upsilon(11020)B_1\bar{B}} \) and \( g_{\Upsilon(11020)B_1\bar{B}^{*}} \) are directly proportional to the corresponding molecular components~\cite{Dong:2009uf}.

\begin{figure}[h!]
\begin{center}
\includegraphics[bb=-20 150 650 415, clip, scale=0.57]{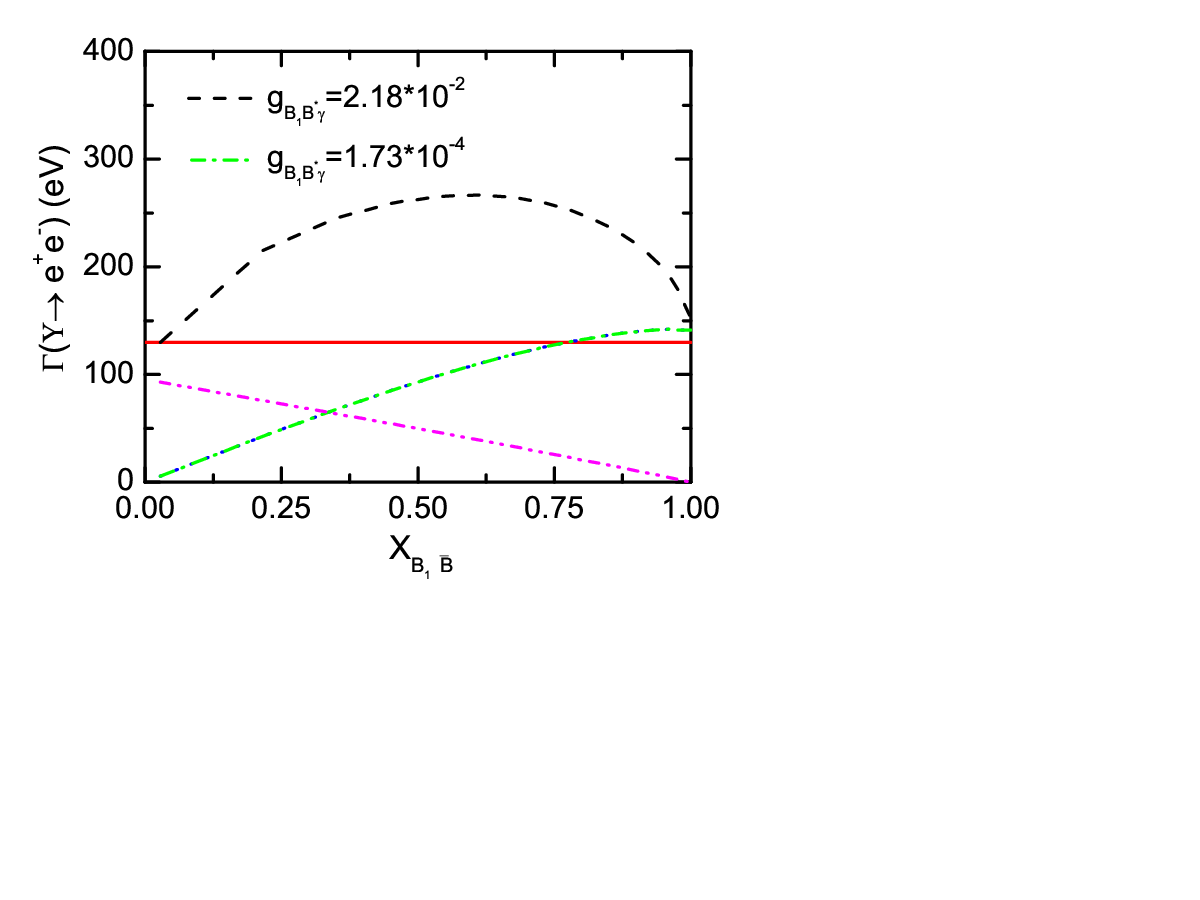}
\caption{Decay width of $\Upsilon(11020)\!\to e^{+}e^{-}$ as a function of the coupling constant $g_{B_1B^{*}\gamma}$ and the $B_1\bar{B}$ molecular component. The red solid line represents the experimental central value of $\Upsilon(11020)\!\to e^{+}e^{-}$, while the black dashed and green dash-dotted lines correspond to $g_{B_1B^{*}\gamma}=0.0218$ and $0.000173$, respectively.}
\label{cc4}
\end{center}
\end{figure}

In our fit to the $\Upsilon(11020)\to e^{+}e^{-}$ decay, a similar opposite trend is observed, as shown by the blue dotted line (overlapping with the green dashed line) and the magenta dash-dotted line in Fig.~\ref{cc4}. This can also be understood as arising from the contributions of the $B_1\bar{B}$ and $B_1\bar{B}^{*}$ molecular components, respectively.
The magenta dash-dotted line corresponds to the coupling constant $g_{B_1B^{*}\gamma}=0.0218$. In this case, the calculated total width can be compared with the experimental value at $\Lambda=0.54$ and $X_{B_1\bar{B}}=0.0279$, indicated by the black dashed line in Fig.~\ref{cc4}. The experimental central value alone is shown as the red solid line.
When the coupling constant is further reduced to $g_{B_1B^{*}\gamma}=0.000173$, the resulting total width becomes nearly identical to that obtained by considering only the $B_1\bar{B}$ component. We also find that for $\Lambda = 0.615$ and $X_{B_1\bar{B}} = 0.761$, the calculated decay width begins to exceed the measured value, providing a further constraint on our model parameters and molecular composition.  Therefore, we use $\Lambda = 0.54$--$0.615$ GeV and $X_{B_1\bar{B}} = 0.0279$--$0.761$ to calculate the decay width.

\begin{table}[h!]
\centering
\tabcolsep=0.6mm
\caption{Partial decay widths of the bottom meson $B_{1}(5721)^0 \to B^{*}\pi$ and $B\gamma{}$.}
\label{tab:cc2}
\begin{tabular}{ccccc}
\hline\hline
State & $J^P$ & Decay channel & Decay width (MeV) & Exp. (MeV) \\
\hline
$B_{1}(5721)^0$ & $1^{+}$ & $B^{*}\pi^{+}$ & $74.83\, g^2_{TH}/\Lambda^2_{\chi}$ & \\
                &         & $B^{*}\pi^{0}$ & $38.25\, g^2_{TH}/\Lambda^2_{\chi}$ & \\
                &         & $B\gamma$      & 0.721 & \\
                &         & $B^{*}\gamma$  & 0.0233 & \\
                &         & Total           & $113.09\, g^2_{TH}/\Lambda^2_{\chi}$+0.721 & $27.5 \pm 3.4$~\cite{ParticleDataGroup:2024cfk} \\
\hline\hline
\end{tabular}
\end{table}
Before presenting the results, we finally need to determine the coupling constant $g_{TH}/\Lambda_{\chi}$. Using the effective Lagrangian in Eq.~(\ref{eq8}),
we compute the partial decay widths of $B_{1}(5721) \to B^{*}\pi$, and the results are summarized in Table~\ref{tab:cc2}. By comparing the calculated total
width with the experimental width of $B_{1}(5721)^0$, we extract $ g_{TH}/\Lambda_{\chi} = 0.487^{+0.030}_{-0.032}$.  Note that the total width reported in
Table~\ref{tab:cc2} does not include the $B_{1} \to B^{*}\gamma$ channel, whose contribution is negligibly small.  Even when the maximal fitted value of the
coupling constant $g_{B_1B^{*}\gamma} = 0.0218$ is inserted into Eq.~\ref{eq7-2}, the corresponding partial width is only $0.0233~\mathrm{MeV}$.

\begin{figure}[h!]
\begin{center}
\includegraphics[bb=-20 100 650 410, clip, scale=0.50]{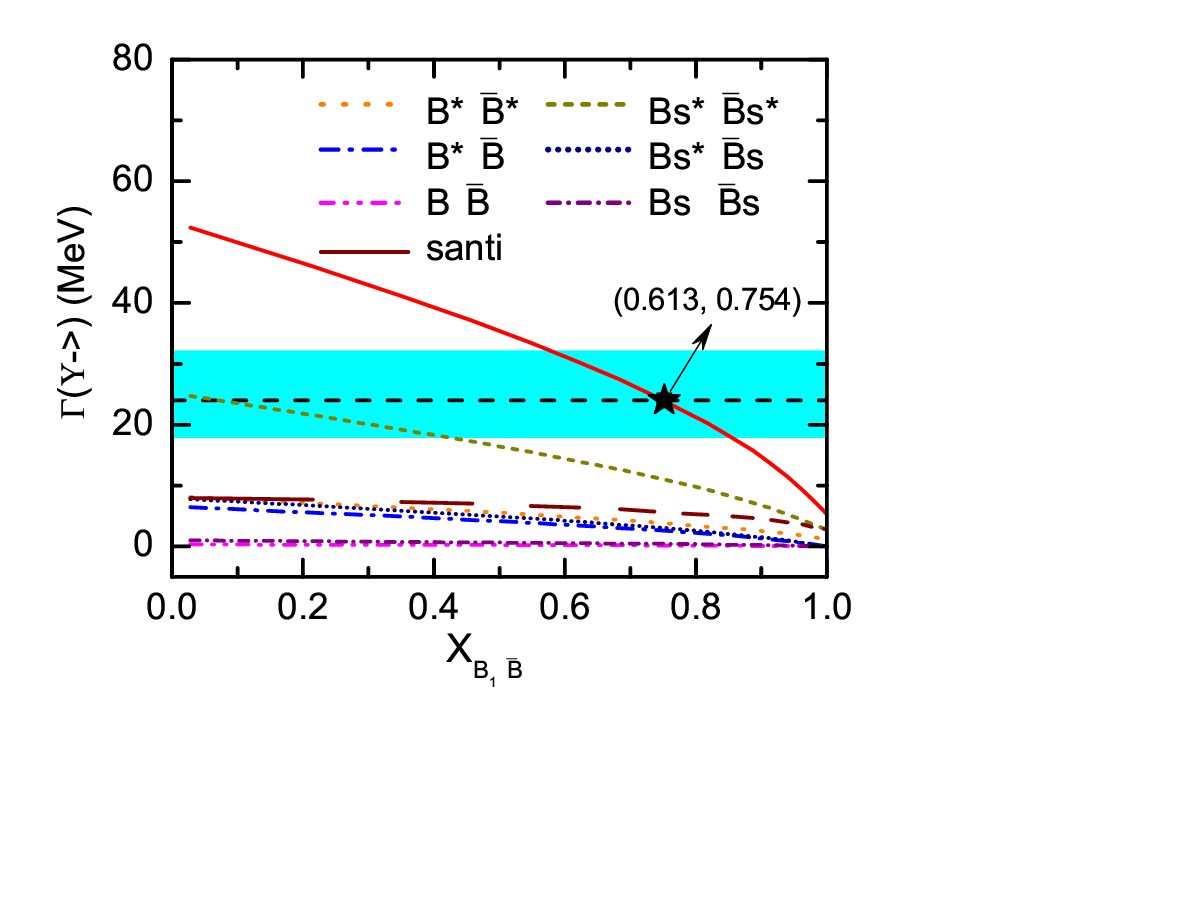}
\caption{Decay widths of $\Upsilon(11020)$ with the red solid line representing the total width, the orange dotted line for $B^{*}\bar{B}^{*}$,
the blue dash-dotted line for $B^{*}\bar{B}$, the magenta dash-dot-dotted line for $B\bar{B}$, the wine two-point-segment solid line for $B^{*}\pi\bar{B}+B^{*}\pi\bar{B}^{*}$,
the dark yellow short-dashed line for $B^{*}_s\bar{B}^{*}_s$, the navy short-dotted line for $B^{*}_s\bar{B}_s$, and the purple short dash-dotted line for $B_s\bar{B}_s$.
The cyan bands show the experimental total width with uncertainties, and the dashed solid line indicates the experimental central value, which matches the theoretical
total width at $\Lambda=0.613$~GeV and $X_{B_1\bar{B}}=0.754$.} \label{cc5}
\end{center}
\end{figure}
With the information obtained above, the total decay widths, including both the two-body and three-body contributions, are presented in Fig.~\ref{cc5} as functions of $X_{B_1\bar{B}}$ in the range $0 \le X_{B_1\bar{B}} \le 1$, where $X_{B_1\bar{B}}=0$ corresponds to the absence of the $B_1\bar{B}$ molecular component and $X_{B_1\bar{B}}=1$ corresponds to a pure $B_1\bar{B}$ molecular state. Only the dominant channels are shown separately, namely the two-body decays
$\Upsilon(11020) \to B^{(*)}\bar{B}^{(*)}$ and $B^{(*)}_{s}\bar{B}_{s}^{(*)}$, as well as the three-body decay $\Upsilon(11020) \to B\pi\bar{B}^{(*)}$.
Contributions from other channels are not displayed individually, but are included in the total width. The cyan bands indicate the experimental total width with
uncertainties, and the dashed solid line represents the experimental central value. From the figure, it is evident that the theoretically calculated total width
decreases monotonically with increasing $B_1\bar{B}$ component.  The results are highly sensitive to the model parameter $\Lambda$ and the molecular component
$X_{B_1\bar{B}}$, with the total width ranging from 52.42~MeV at $\Lambda=0.54$~GeV and $X_{B_1\bar{B}}=0.027$ to 3.85~MeV at $\Lambda=0.707$~GeV and $X_{B_1\bar{B}}=1$.

Using the fitted values of the model parameter $\Lambda$ and the molecular component $X_{B_1\bar{B}}$, which vary consistently within the ranges
$\Lambda = 0.54$--$0.615$~GeV and $X_{B_1\bar{B}} = 0.0279$--$0.761$ (as summarized in Table~\ref{table2}), the calculated total decay widths of the
$\Upsilon(11020)$ are in good agreement with the experimental measurements over the entire parameter range.
In particular, for $\Lambda = 0.613$~GeV, where the $\Upsilon(11020)$ is assumed to contain a $75.4\%$ $B_1\bar{B}$ molecular component,
the calculated total width coincides with the experimental central value of $24$~MeV, as indicated by the pentagram in Fig.~\ref{cc5}.
This agreement demonstrates that the experimentally measured total decay width can be well reproduced within our framework,
providing strong and direct evidence that the $\Upsilon(11020)$ can be interpreted as an $S$-wave $B_1\bar{B}$--$B_1\bar{B}^{*}$ molecular state,
dominated by the $B_1\bar{B}$ component.

Our results suggest that, when the $\Upsilon(11020)$ is interpreted as a $B_1\bar{B}$--$B_1\bar{B}^{*}$ molecular state, its dominant decay channel is
$B_s^{*}\bar{B}_s^{*}$, which has not been observed experimentally, with a partial width of $10.573$~MeV, larger than that of the three-body decay channel
$B^{*}\pi\bar{B}+B^{*}\pi\bar{B}^{*}$, also not seen experimentally, which is induced by the tree-level contribution and calculated to be $5.573$~MeV. This behavior stands in
stark contrast to interpretations of the $\Upsilon(11020)$ as a conventional $b\bar{b}$ quarkonium state; for instance, Ref.~\cite{Godfrey:2015dia} found
that the $B\bar{B}$ decay channel accounts for $58.69\%$ of the total width, whereas the $B_s^{*}\bar{B}_s^{*}$ channel contributes only $0.914\%$. Furthermore,
an analysis of $e^+e^-$ annihilation data into bottomonium states~\cite{Dong:2020tdw,Belle:2021lzm} using the K-matrix method indicates that the dominant
decay channel of the $\Upsilon(11020)$ is $B_s\bar{B}_s$, accounting for approximately $70\%$--$90\%$ of the total width~\cite{Husken:2022yik}, which is
significantly larger than the $B_s\bar{B}_s$ partial width obtained in our calculation, amounting to only $0.418$~MeV at $\Lambda = 0.615$~GeV. In the
framework of the non-relativistic quark model, the dominant decay channel of the $\Upsilon(11020)$ is predicted to be $B^{*}\bar{B}^{*}$, with a branching
fraction exceeding $50\%$~\cite{Hapareer:2023yvl}, in agreement with calculations performed using the $^{3}P_0$ model~\cite{Ferretti:2013vua,Segovia:2016xqb},
in contrast to our calculation, where this channel contributes only a small fraction of the total width. Taken together, these differences in the decay patterns
could serve as crucial experimental signatures for testing whether the $\Upsilon(11020)$ can be interpreted as a $B_1\bar{B}$--$B_1\bar{B}^{*}$ molecular state.

\begin{table}[h!]
\centering
\tabcolsep=3.3mm
\caption{Partial decay widths of the $\Upsilon(11020)$ into experimentally observed channels.}
\label{tab:cc3}
\begin{tabular}{cc|ccc}
\hline\hline
  Channel          & Width                &  Channel              & Width \\
\hline
  $\pi\pi\Upsilon(1S)$    & $8.242$  (eV)          &$\pi\pi\pi\chi_{b0}$   & 0.754 (KeV) \\
  $\pi\pi\Upsilon(2S)$    & $29.715$  (eV)          &$\pi\pi\pi\chi_{b1}$   &0.167 (MeV) \\
  $\pi\pi\Upsilon(3S)$    & $161.592$  (eV)        &$\pi\pi\pi\chi_{b2}$   &0.048 (MeV)\\
  $\pi\pi{}h_b(1P)$       & $37.281$ (eV)          &$\pi\pi{}h_b(2P)$   &0.117 (KeV)\\
\hline\hline
\end{tabular}
\end{table}
Finally, under the interpretation of $\Upsilon(11020)$ as a $B_1\bar{B}$--$B_1\bar{B}^{*}$ molecular state, Table~\ref{tab:cc3} presents the partial decay widths
for experimentally observed channels, calculated at $\Lambda = 0.613$~GeV with $X_{B_1\bar{B}} = 0.754$.  The decays $\Upsilon(11020) \to \pi\pi \Upsilon(nS)$ and
$\Upsilon(11020) \to \pi\pi h_b(nP)$ proceed via the intermediate resonances $Z_b$ and $Z_b^{\prime}$, as illustrated in Fig.~\ref{cc1}(c--d), with partial widths
on the order of a few eV. This result differs significantly from predictions based on conventional quark models, which generally yield partial widths on the order of keV~\cite{Segovia:2016xqb}. Through the intermediate $\omega$ meson, the decay width of the channel $\Upsilon(11020) \to \pi\pi\pi\chi_{bJ}$ increases substantially,
with the $\pi\pi\pi\chi_{b1}$ final state reaching 0.167~MeV, while the smallest $\pi\pi\pi\chi_{b0}$ channel, which has not yet been observed experimentally,
remains 0.754~keV in our model.

\section{Summary}\label{sec:summary}
In this work, we investigate whether the experimentally observed $\Upsilon(11020)$ can be interpreted as an $S$-wave $B_1\bar{B}$--$B_1\bar{B}^{*}$ molecular state. Using the compositeness condition and effective Lagrangians, we calculate the strong decay widths of $\Upsilon(11020)$ within the molecular framework. By fitting existing experimental data, including $\Upsilon(11020)\to e^+ e^-$ and $\Upsilon(11020)\to \chi_{bJ}\pi\pi\pi$ decays, we extract the couplings of $\Upsilon(11020)$ to its constituents $B_1$ and $\bar{B}^{(*)}$.

With the extracted coupling, we calculate the partial decay widths of the $\Upsilon(11020)$ into $B^{(*)}_{(s)}\bar{B}^{(*)}_{(s)}$, $\pi\pi \Upsilon(nS)$, $\pi\pi h_b(nP)$, and $\pi\pi\pi \chi_{b1}$ via hadronic loops, as well as the three-body $B^{*}\pi \bar{B}^{(*)}$ decays through tree-level diagrams (see Fig.~\ref{cc1}). The results indicate that the $\Upsilon(11020)$ is predominantly a $B_1\bar{B}$ molecular state, accounting for approximately 75.4\% of its total composition.  Its primary decay channel is the strange $B_s^{*}\bar{B}^{*}_s$ mode, which has not yet been observed experimentally.  Notably, the partial widths for $\pi\pi \Upsilon(nS)$ and $\pi\pi h_b(nP)$ are only a few eV: $8.242$, $29.715$, and $161.592$~eV for $\pi\pi \Upsilon(1S)$, $\pi\pi \Upsilon(2S)$, and $\pi\pi \Upsilon(3S)$, respectively, and $37.281$~eV and $0.117$~keV for $\pi\pi h_b(1P)$ and $\pi\pi h_b(2P)$. In contrast, the $\pi\pi\pi \chi_{bJ}$ channels, especially $\pi\pi\pi \chi_{b1}$, are significantly enhanced, reaching up to $0.167$~MeV.
The yet-unobserved $\pi\pi\pi \chi_{b0}$ channel could be as large as 0.754~keV. These features differ substantially from the predictions of conventional quark models and provide a clear experimental signature for the molecular interpretation of $\Upsilon(11020)$.

Such distinctive decay patterns could be probed at LHCb and other experimental facilities. Once experimentally confirmed, they would further deepen our understanding of heavy-quark symmetry.

\section*{Acknowledgments}
This work was supported by the Sailing Plan Project of Yibin University (No. 2021QH06).  Y. Huang also acknowledges support from the National Natural Science Foundation of China under Grant No. 12005177.

\appendix
\section{Decay Width Formula for $\Upsilon(11020)\to \chi_{bJ}\pi\pi\pi$}
\label{app:decay_width}
In the narrow-width approximation (NWA), $\Gamma_X \ll M_X$ (valid for the $\omega$ meson)
and neglecting vertex dependence on the virtuality of $X$, the cascade decay factorizes.
For the chain
\begin{align}
\Upsilon(11020)\to \chi_{bJ}\, \omega \to \chi_{bJ}\, \pi\pi\pi,
\end{align}
let $A\equiv \Upsilon(11020)$, $X\equiv \omega$ (mass $M_X$, width $\Gamma_X$), $B\equiv \chi_{bJ}$,
and $n$-body final state ($n=3$) with invariant mass $M_n^2 \equiv (p_{\pi_1}+p_{\pi_2}+p_{\pi_3})^2$.
The total amplitude via $X$ reads
\begin{align}
\mathcal{M}_{A\to B+1\cdots n}&= \mathcal{M}_{A\to B X}(M_n) \frac{1}{M_n^2-M_X^2+iM_X\Gamma_X}\nonumber\\
                              &\times\mathcal{M}_{X\to 1\cdots n}(M_n),
\end{align}
and the $(n+1)$-body width is
\begin{align}
\Gamma_{A\to B,1\cdots n}
&= \frac{1}{2M_A} \int d\Phi_{n+1} \sum_{\rm spins}|\mathcal{M}_{A\to B,1\cdots n}|^2.
\end{align}

Decomposing the phase space,
\begin{align}
d\Phi_{n+1} &= d\Phi_2(A\to B X)\frac{dM_n^2}{2\pi} d\Phi_n(X\to 1\cdots n),
\end{align}
yields
\begin{align}
\frac{d\Gamma_{A\to B,1\cdots n}}{dM_n^2}
&= \Gamma_{A\to B X}(M_n)\, \frac{1}{\pi} \frac{M_n \Gamma_{X\to 1\cdots n}(M_n)}{(M_n^2-M_X^2)^2 + M_X^2 \Gamma_X^2}.
\end{align}

Changing variables to $M_n$ gives
\begin{align}
\frac{d\Gamma_{A\to B,1\cdots n}}{dM_n}
&= \Gamma_{A\to B X}(M_n)\, \frac{2 M_n^2}{\pi} \frac{\Gamma_{X\to 1\cdots n}(M_n)}{(M_n^2-M_X^2)^2 + M_X^2 \Gamma_X^2}.
\end{align}

Under the NWA and assuming weak $M_n$ dependence, evaluating at $M_n=M_X$ and using
\begin{align}
\int \frac{dM_n^2}{\pi} \frac{M_X \Gamma_X}{(M_n^2-M_X^2)^2 + M_X^2 \Gamma_X^2} &= 1,
\end{align}
gives the factorized width
\begin{align}
\Gamma(A\to B,1\cdots n)
&\approx \Gamma(A\to B X) \frac{\Gamma(X\to 1\cdots n)}{\Gamma_X} \nonumber\\
&= \Gamma(A\to B X)\, {\rm Br}(X\to 1\cdots n),
\end{align}
and for the specific case,
\begin{align}
\Gamma(\Upsilon(11020)\to \chi_{bJ}\pi\pi\pi)
&\approx \Gamma(\Upsilon(11020)\to \chi_{bJ}\omega)\, {\rm Br}(\omega\to 3\pi).
\end{align}


%
\end{document}